\pdfoutput=1

\documentclass[prl,superscriptaddress,amsmath,amssymb,showpacs,twocolumn]{revtex4-1}

\usepackage{graphicx,bm,amsmath}
\usepackage[usenames,dvipsnames]{color}
\usepackage[colorlinks,bookmarks=false,citecolor=NavyBlue,linkcolor=Red,urlcolor=blue]{hyperref}
\usepackage[bbgreekl]{mathbbol}

\newcommand{\be}{\begin{equation}}
\newcommand{\ee}{\end{equation}}
\newcommand{\bea}{\begin{eqnarray}}
\newcommand{\eea}{\end{eqnarray}}

\newcommand{\up}{\uparrow}
\newcommand{\down}{\downarrow}
\newcommand{\neel}{|\cdots\up\down\up\down\cdots\rangle}
\newcommand{\aneel}{|\cdots\down\up\down\up\cdots\rangle}

\def\nn{\nonumber\\}
\def\fr#1{(\ref{#1})}
\def\doi{http://dx.doi.org/}

\begin{document}
\title{How order melts after quantum quenches}
\author{Mario Collura}
\affiliation{Theoretische Physik, Universit\"at des Saarlandes, D-66123 Saarbr\"ucken, Germany.}
\affiliation{Dipartimento di Fisica e Astronomia ``G. Galilei'', Universit\`a di Padova, I-35131 Padova, Italy}   
\author{Fabian H.L. Essler}
\affiliation{Rudolf Peierls Centre for Theoretical Physics,
Oxford University, Oxford, OX1 3PU, United Kingdom}
    
\begin{abstract}
Injecting a sufficiently large energy density into an isolated
many-particle system prepared in a state with long-range order will
lead to the melting of the order over time. Detailed
information about this process can be derived from the quantum
mechanical  probability distribution of the order parameter. We
study this process for the paradigmatic case of the spin-1/2
Heisenberg XXZ chain. We determine the full quantum 
mechanical distribution function of the staggered subsystem
magnetization as a function of time after a quantum quench from the
classical N\'eel state. This provides a detailed picture
of how the N\'eel order melts and reveals the existence of an
interesting regime at intermediate times that is characterized by a
very broad probability distribution.
\end{abstract}
\maketitle
\paragraph{Introduction. ---}
A fundamental objective of quantum theory is to determine probability
distribution functions of observables in given quantum states. In
few-particle systems the time evolution of such probability
distributions provides a lot of useful information beyond what is
contained in the corresponding expectation values.
Recent advances in cold-atom experiments have made possible not only
the study of non-equilibrium time evolution of (almost) isolated
many-particle systems~\cite{Langen2015,Altman2015,kww-06,GM:col_rev02,hacker10,tetal-11,shr-12,cetal-12,langen13,MM:Ising13,ronzheimer13,zoran1,lev2017},
but given access to the full quantum mechanical probability
distributions of certain observables~\cite{HLSI08,KPIS10,KISD11,GKLK12,Greiner}.  
This provides an opportunity to gain new insights about the coherent
dynamics of many-particle quantum systems. One intriguing
question one may ask is how order melts, or forms, when an 
isolated many-particle system is driven across a phase
transition. Related questions have been studied in solids, but there
one essentially deals with open quantum systems and has access to very
different observables, see e.g.~\cite{Hill,Cavalieri}. 
The basic setup we have in mind is as follows. Let us consider a
system of quantum spins with Hamiltonian $H$ that is initially
prepared in a state with density matrix $\rho(0)$. In this state there
is long-range order characterized by a local order parameter
${\cal O}=\sum_{j=1}^L {\cal O}_{j}$, where $j$ runs over the sites of
the lattice and ${\cal O}_j$ is a local operator. We are interested in
the probability distribution $P_A$ of the order parameter ${\cal O}_A$
in a contiguous subsystem of linear size $|A|$
\be
P_{A}(m,t)={\rm Tr}\big[\rho(t)\delta({\cal O}_A-m)\big].
\ee
Here $\rho(t)$ is the density matrix of the system at time $t$ and
$P_{A}(m,t)$ is the probability that the subsystem order parameter
${\cal O}_A$ takes the value $m$ in the state $\rho(t)$. We are
interested in cases where the system is initially well ordered at all
length scales and $P_{A}(m,t)$ is therefore narrowly peaked around the
average ${\cal O}_A$. Under time evolution the order melts and
at late times and large subsystem sizes $P_{A}(m,t)$ is believed to approach a
Gaussian distribution centred around
zero~\cite{Deutsch1991,Srednicki1994,Rigol2008,Rigol2007,Cassidy2011,Caux2013,Ilievski2015}. 
The question of interest is how $P_{A}(m,t)$ evolves as a function of
time and subsystem size $|A|$. Varying the latter provides information
about how well the system is ordered at length scale $|A|$.

We find that when the initially ordered system is quenched well into the
unbroken symmetry phase of the Hamiltonian, the (local) order quickly
disappears and the PDF acquires a simple Gaussian shape. In contrast,
when the quantum quench is to an energy density where Hamiltonian
eigenstates retain short-range order, the PDF exhibits a complex
structure both at finite times and in the steationay state.
In the following we focus on the example of the spin-$1/2$ Heisenberg XXZ chain, 
but note that the picture we put forward is general and has a wide
range of applicability.

\begin{figure}[t!]
\begin{center}
\includegraphics[width=0.4\textwidth]{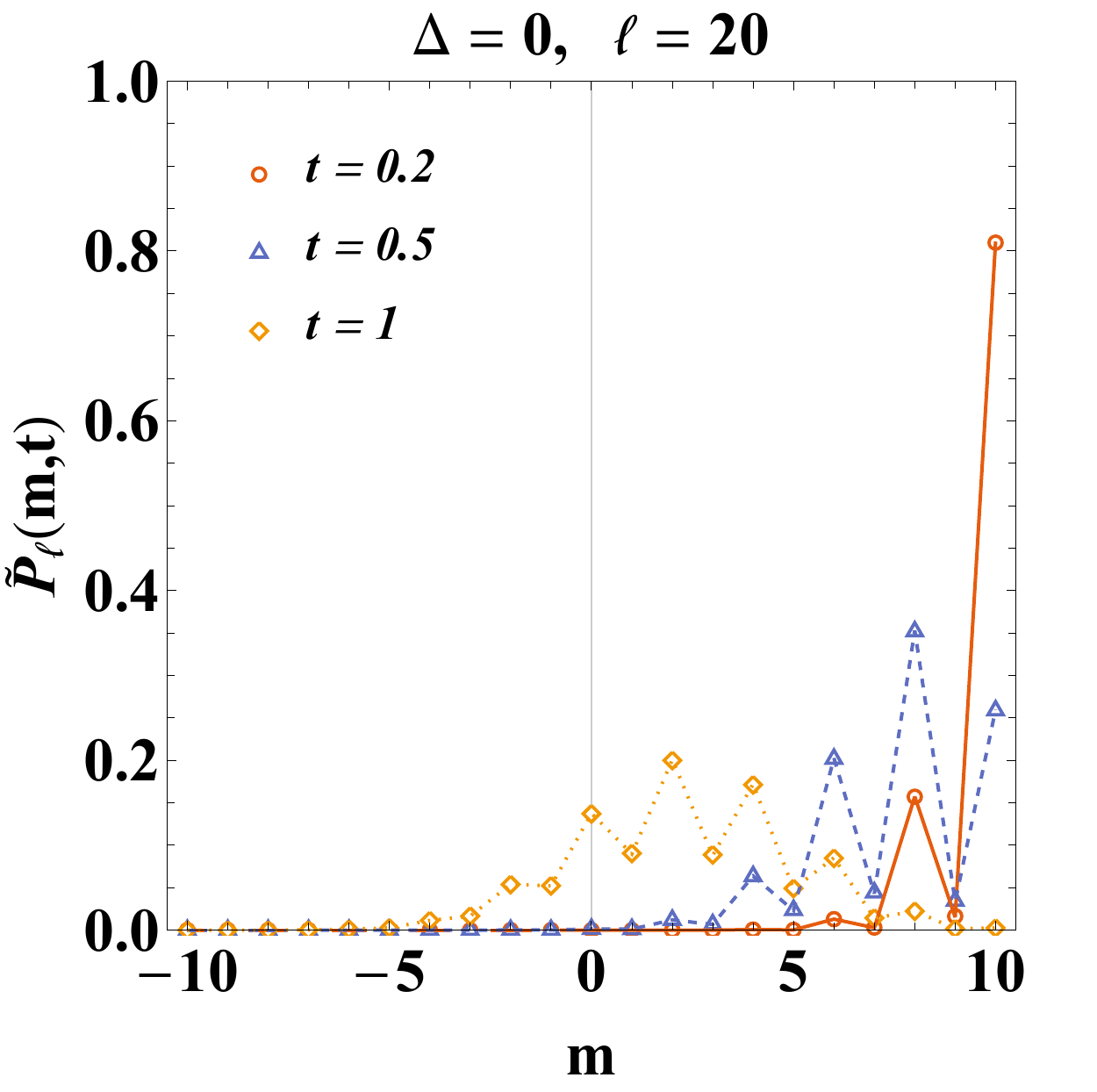}
\caption{$\widetilde P_\ell(m,t)$ for $\ell=20$ at times $t=0,0.5,1$ after a quantum
quench from a classical N\'eel state to a Heisenberg chain with $\Delta=0$.
Lines are exact results (see main text) and symbols are obtained by iTEBD. 
\label{fig:short0}
}
\end{center}
\end{figure}

\begin{figure}[t!]
\begin{center}
\includegraphics[width=0.4\textwidth]{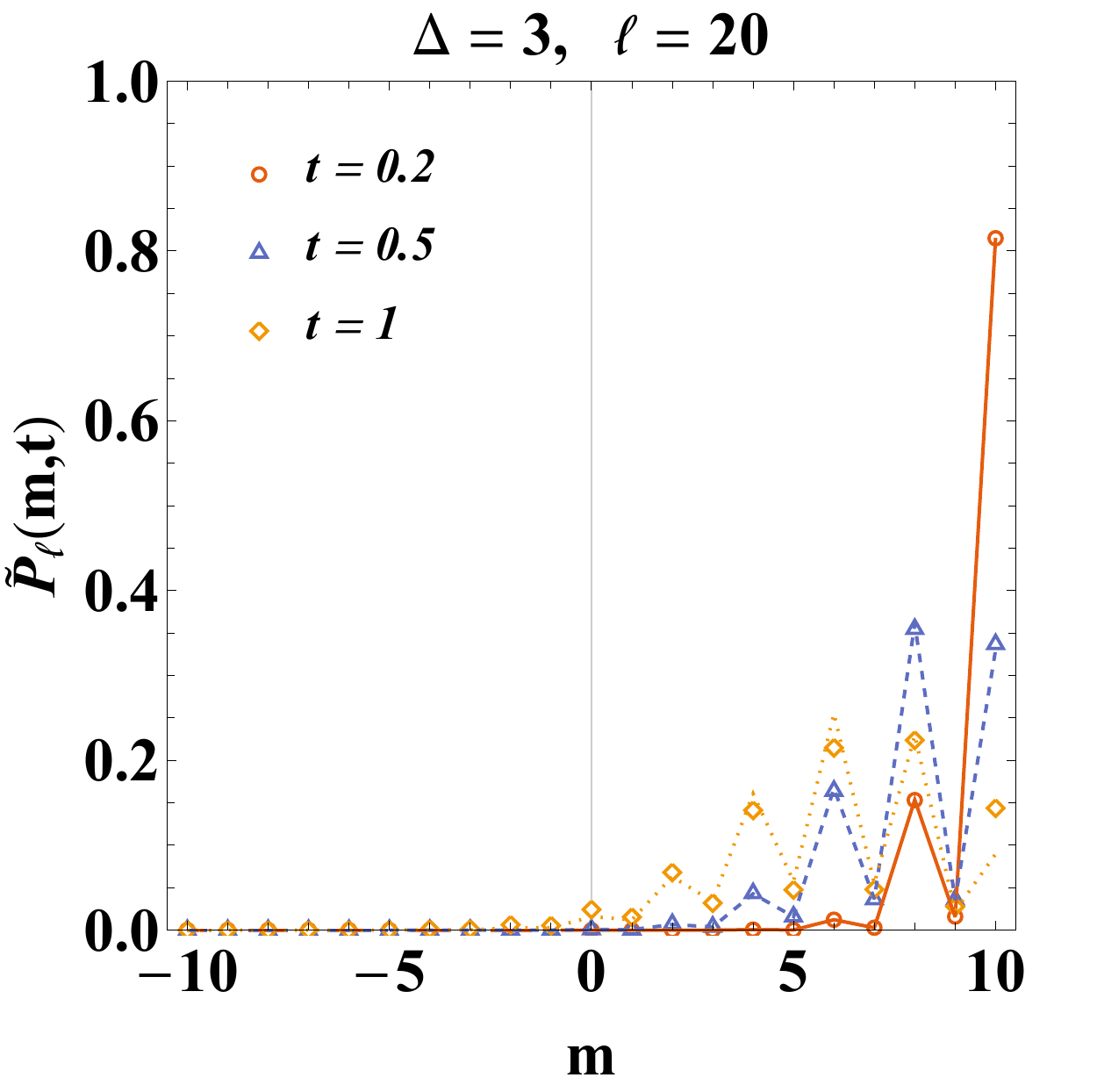}
\caption{$\widetilde P_\ell(m,t)$ for $\ell=20$ at times $t=0$, $0.5$, $1$ after
a quench from the classical N\'eel state to a Heisenberg chain with
$\Delta=4$. Lines are obtained by the self-consistent fermionic
mean-field approximation and the symbols are iTEBD results.
}
\label{fig:short3}
\end{center}
\end{figure}

\paragraph{Model and Setup. ---}
\label{sec:model_setup}
We investigate the time evolution of antiferromagnetic (short-ranged)
order after a quantum quench in the spin-$1/2$ XXZ chain 
\be\label{eq:H}
H_{\Delta} = 
\sum_{j} 
S^{x}_{j}S^{x}_{j+1}
+ S^{y}_{j}S^{y}_{j+1}
+ \Delta \, S^{z}_{j}S^{z}_{j+1} .
\ee
Here $S^{\alpha}_{j}$ are spin-$1/2$ operators acting on the site $j$
and we restrict our analysis to the range $\Delta>0$. The phase
diagram of (\ref{eq:H}) is well established: at $T=0$ there is a BKT
phase transition at $\Delta=1$ that separates a quantum critical phase
at $\Delta<1$ and an antiferromagnetically ordered phase at
$\Delta>1$. At any finite temperature the antiferromagnetic order melts.
The Hamiltonian \fr{eq:H} is invariant under rotations by an arbitrary
angle around the z-axis, translations by one site, and rotations
around the x-axis by 180 degrees. In the thermodynamic limit at
$\Delta>1$ and zero temperature the last symmetry gets broken
spontaneously and one of the two degenerate ground states $|{\rm
GS}^{\pm}_{\Delta}\rangle$, characterized by equal but opposite
expectation values of the staggered magnetization per site, gets selected. 
In the Ising limit $\Delta\to\infty$ the ground states become the
classical N\'eel states, i.e. $|{\rm GS}^{+}_{\infty}\rangle = \neel$
and $|{\rm GS}^{-}_{\infty}\rangle = \aneel$. 
In order to investigate the melting of antiferromagnetic order we
consider the following quantum quench protocol:
(i) we prepare the system in the classical N\'eel state
$|\Phi_{0}\rangle = |{\rm GS}^{+}_{\infty}\rangle$, which exhibits
saturated antiferromagnetic long-ranged order;
(ii) We consider unitary time evolution with Hamiltonian
$H_{\Delta}$. The state of the system at time $t$ is thus
$|\Phi_{t}\rangle = \exp(-i H_{\Delta} t) |\Phi_{0}\rangle$.
This quench is integrable~\cite{Pozsgay:13a,FE_13b,FCEC_14} and exact
results on the stationary state are
available~\cite{neel_overlap,QANeel,XXZung,XXZunglong,Ilievski2015,IQNB}. 
We employ the infinite Time-Evolving Block-Decimation (iTEBD)
algorithm~\cite{iTEBD1,iTEBD2} to obtain a very accurate description
of $|\Phi_{t}\rangle$ in the thermodynamic limit. However, the growth
of the bipartite entanglement entropy limits the time window
accessible by this method. Retaining up to $\chi_{max} = 1024$
auxiliary states, we are able to reach a time $t_{max} \simeq 12$
without significant error ($\lesssim 10^{-3}$). 
\begin{figure}[t!]
\begin{center}
\includegraphics[width=0.225\textwidth]{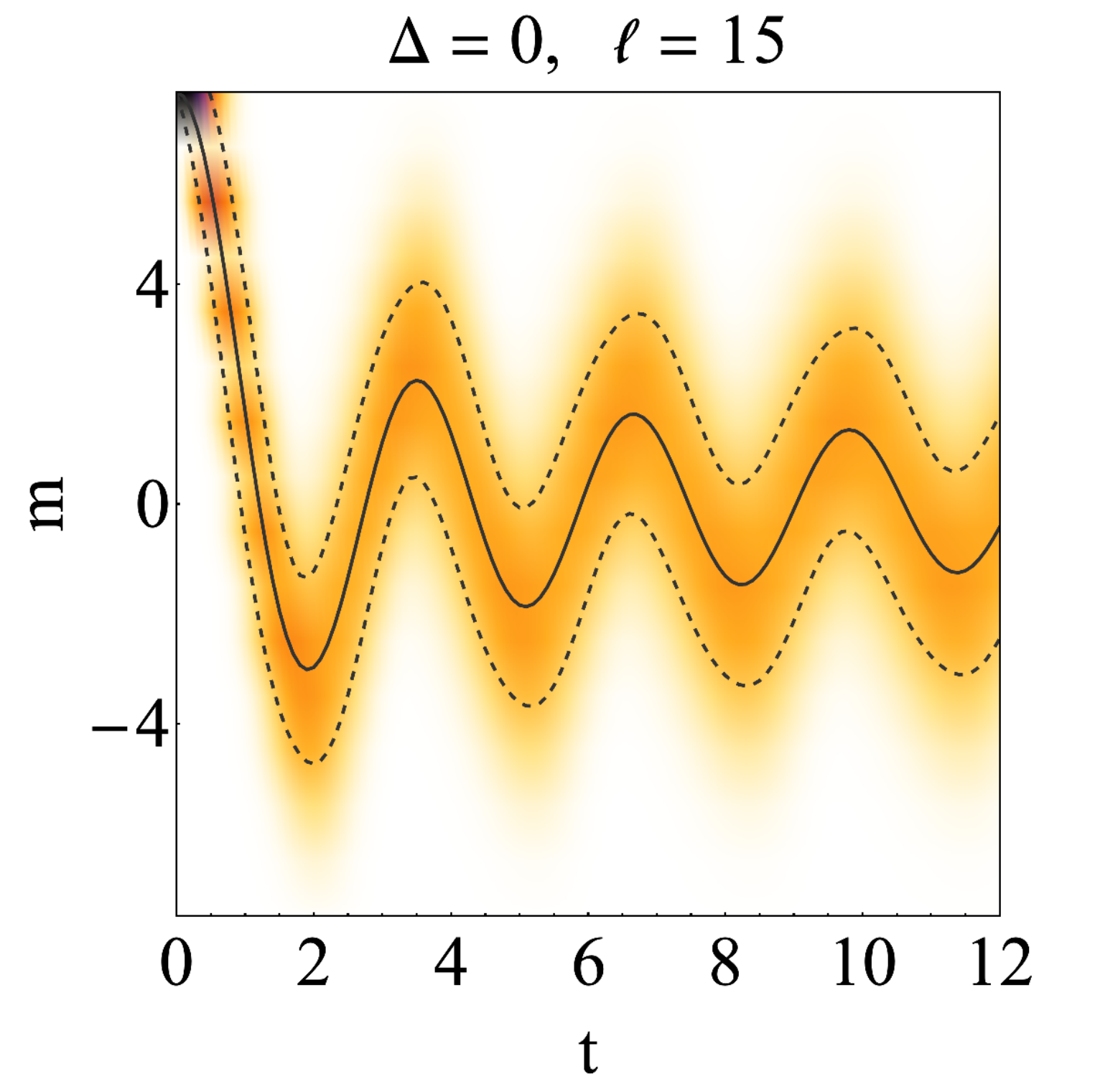}
\includegraphics[width=0.225\textwidth]{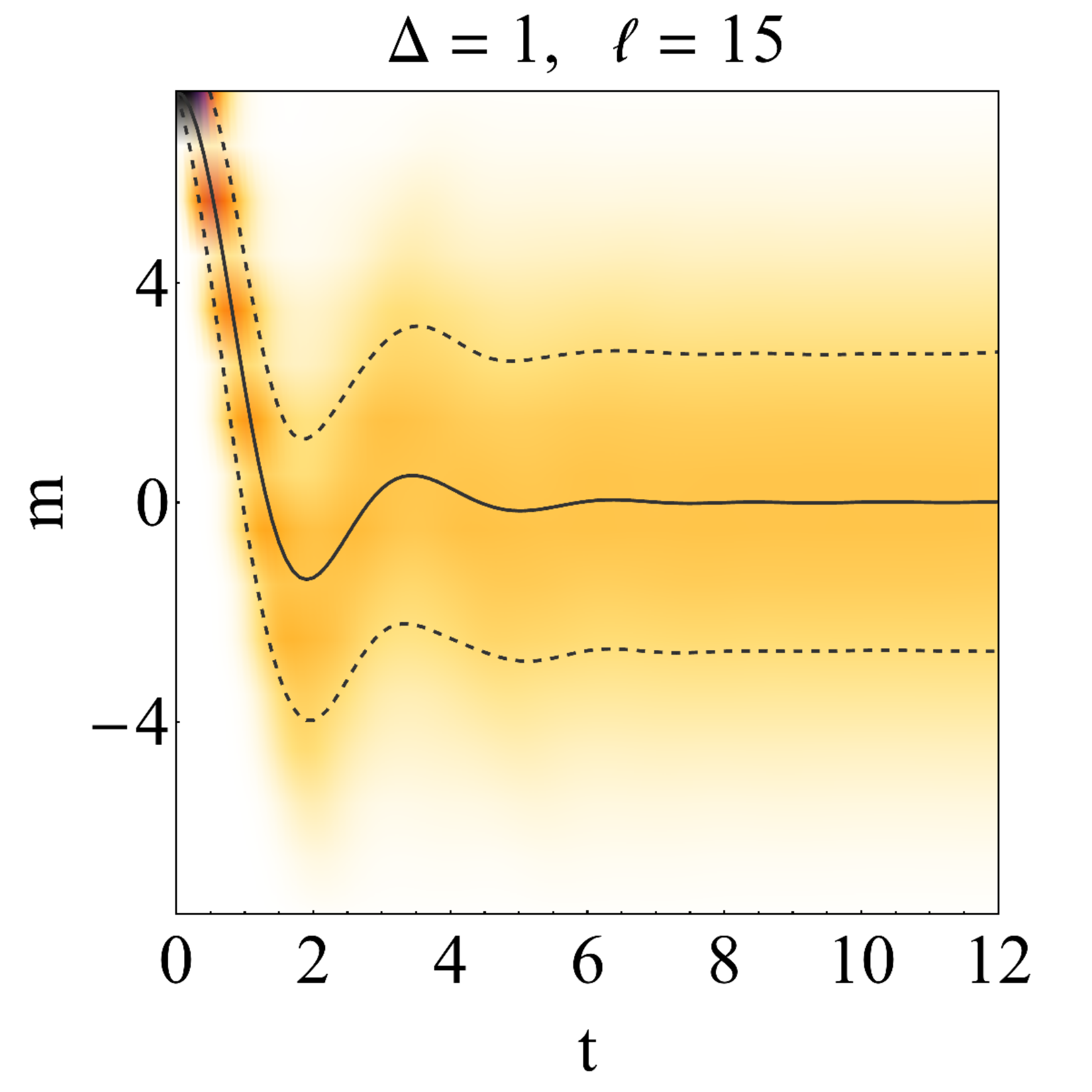}
\raisebox{14pt}{\includegraphics[height=0.18\textwidth]{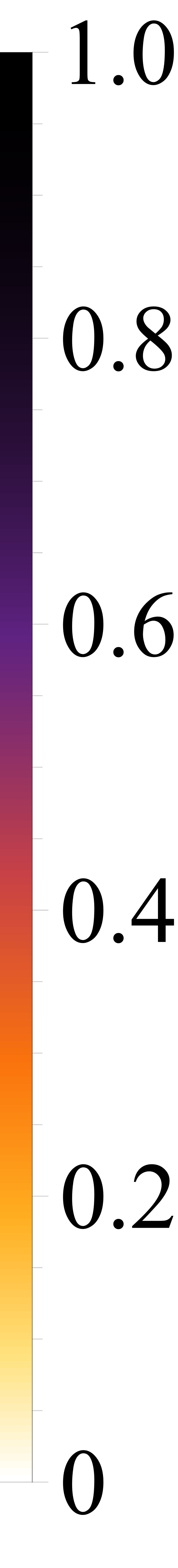}}\\
\includegraphics[width=0.225\textwidth]{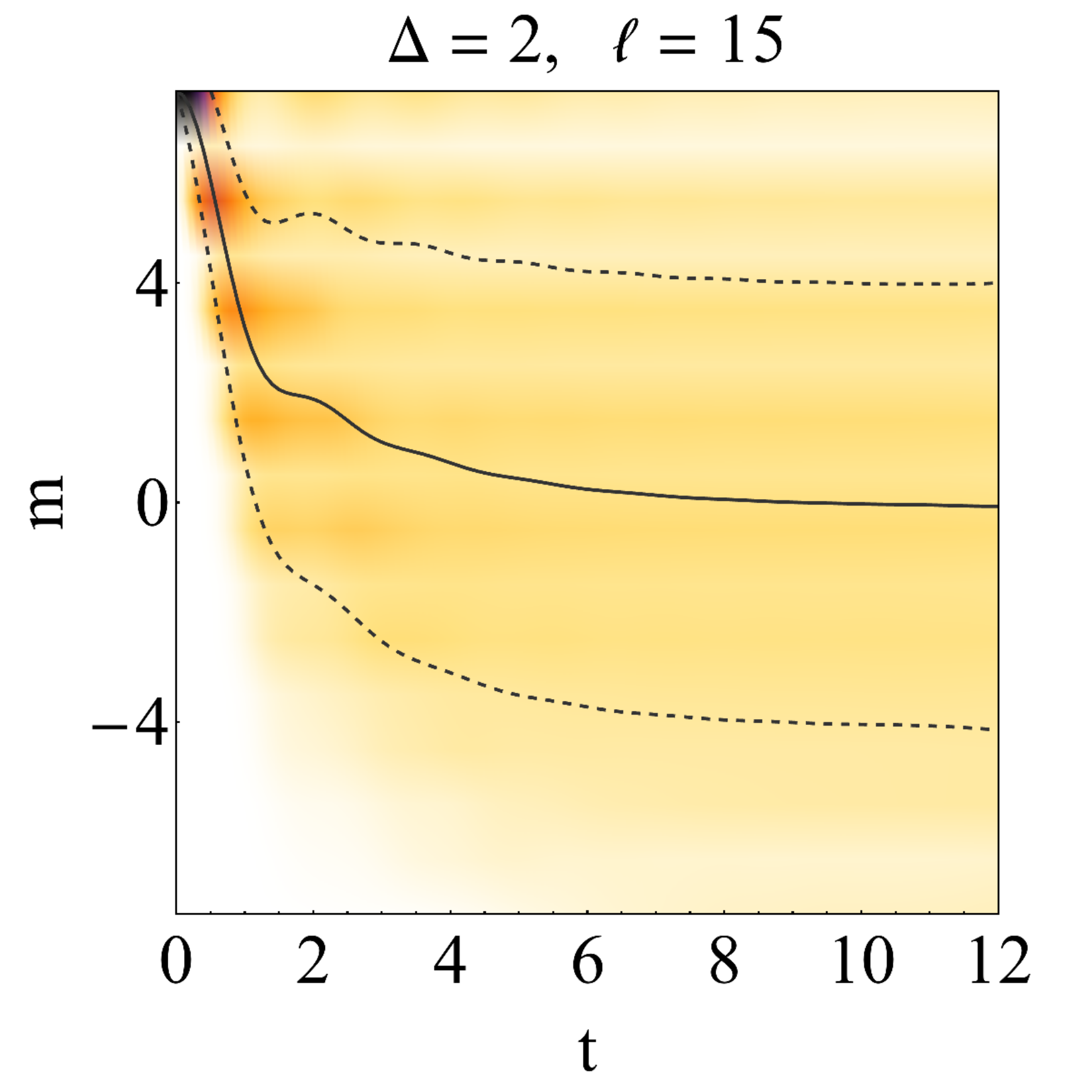}
\includegraphics[width=0.225\textwidth]{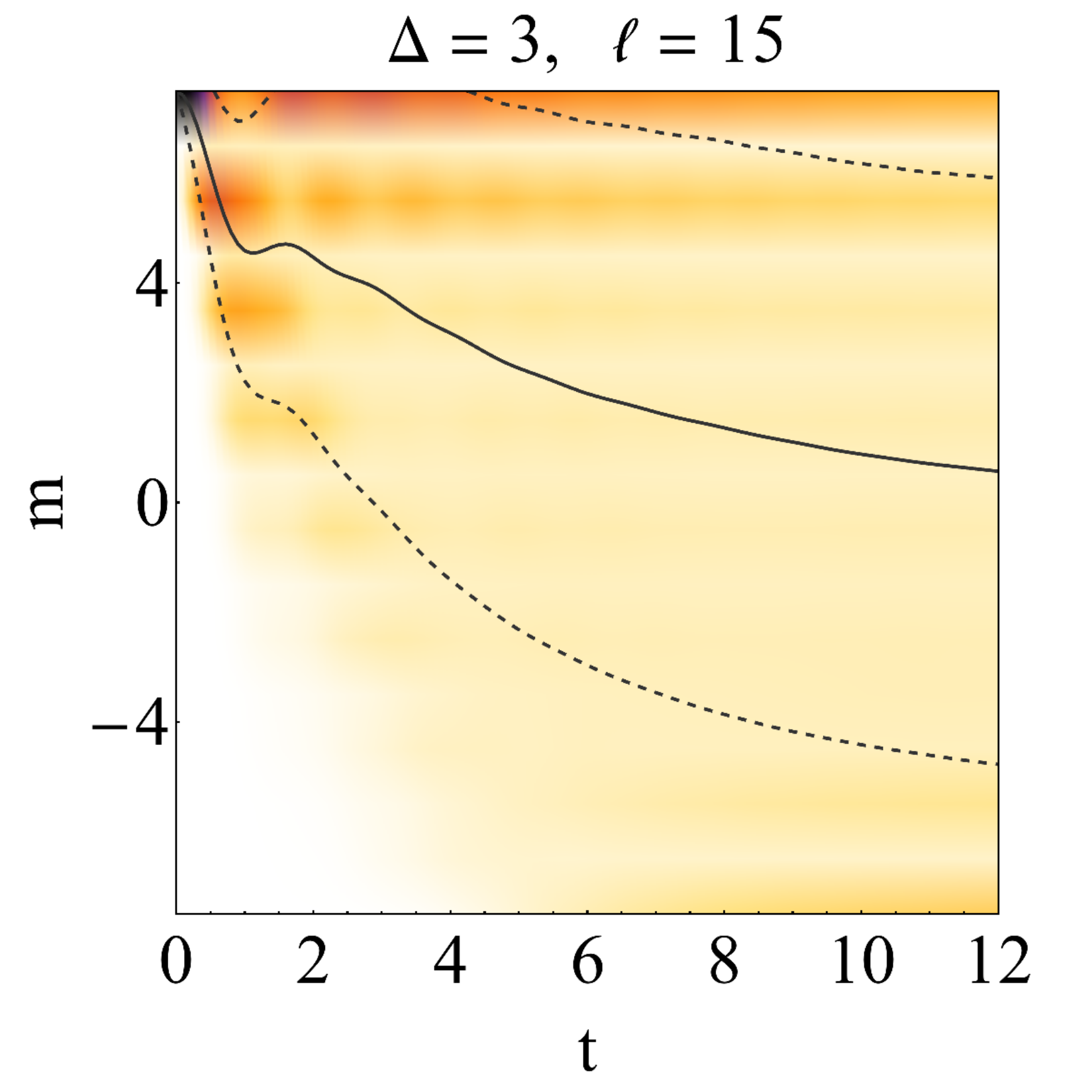}
\raisebox{14pt}{\includegraphics[height=0.18\textwidth]{bar_r.pdf}}
\caption{\label{fig:P_density}
Density plot of $\widetilde P_{\ell}(m, t)$ with
$m\in[-\ell/2,\ell/2]$ and $t\in[0,12]$ for a subsystem size $\ell =
15$, after a quench from the N\'eel state  $|{\rm
GS}^{+}_{\infty}\rangle$ to $\Delta = 0,\, 1,\, 2, \, 3$. 
The full line represents the expectation value $\bar{m}(t)$, 
the dashed lines are the standard deviation from the average, namely
$\bar{m}(t) \pm  \sigma(t)$.
}
\end{center}
\end{figure}
\paragraph{PDF dynamics. ---}\label{sec:PDF}
Detailed information on how the antiferromagnetic order melts as the system
evolves in time is provided by the probability distribution of the
staggered subsystem magnetisation $M_{\ell} \equiv \sum_{j = 1}^{\ell}
(-1)^{j} S^{z}_{j}$
\bea\label{eq:P}
P_{\ell}(m, t) &\equiv&  \langle\Phi_t| \delta(M_{\ell}  -
m)|\Phi(t) \rangle\ ,\nn
&=& \sum_{r\in\mathbb{Z}}\widetilde P_{\ell}(m,
t)  \delta\big(m - r -(1- (-1)^{\ell})/4\big)\ ,
\eea
where second line follows from the fact that the eigenvalues of
$M_\ell$ are half-integer numbers. We note that the probabilities satisfy the
normalisation condition $\sum_{m = -\ell/2}^{\ell/2} \widetilde
P_{\ell}(m, t) = 1$. The initial N\'eel state is an eigenstate of the
staggered subsystem magnetization $M_\ell$ and concomitantly the probability
distribution is a delta function $P_{\ell}(m,
0)=\delta(m-\ell/2)$. This reflects the long-range magnetic order in
the initial state. In Fig.~\ref{fig:P_density} we show the evolution
of $P_\ell(m,t)$ in time obtained by iTEBD for subsystem size
$\ell=15$ and several values of the interaction strength $\Delta$ in
the ``post-quench'' Hamiltonian \fr{eq:H}.
\begin{figure}[t!]
\begin{center}
\includegraphics[width=0.2325\textwidth]{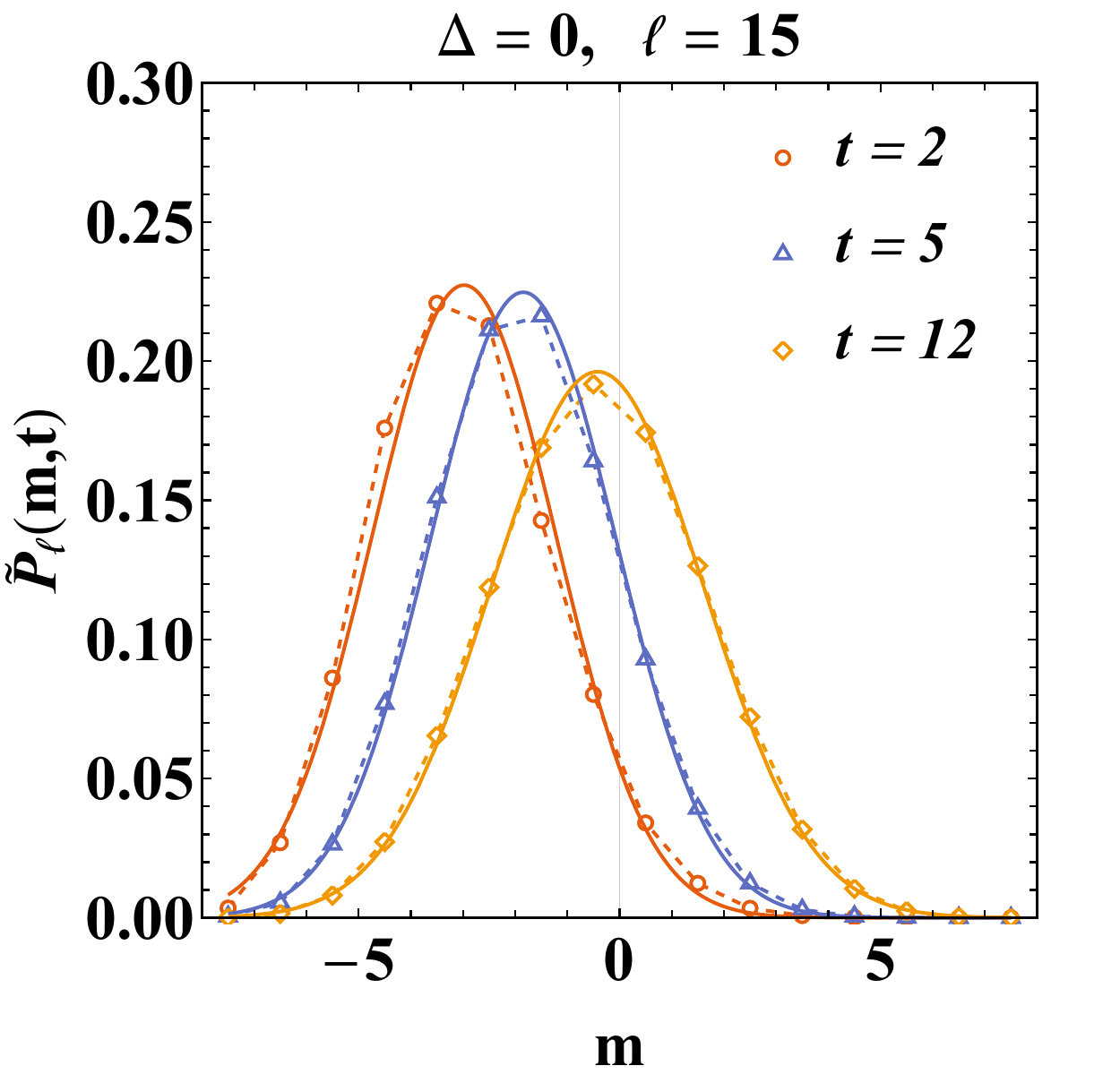}
\includegraphics[width=0.2325\textwidth]{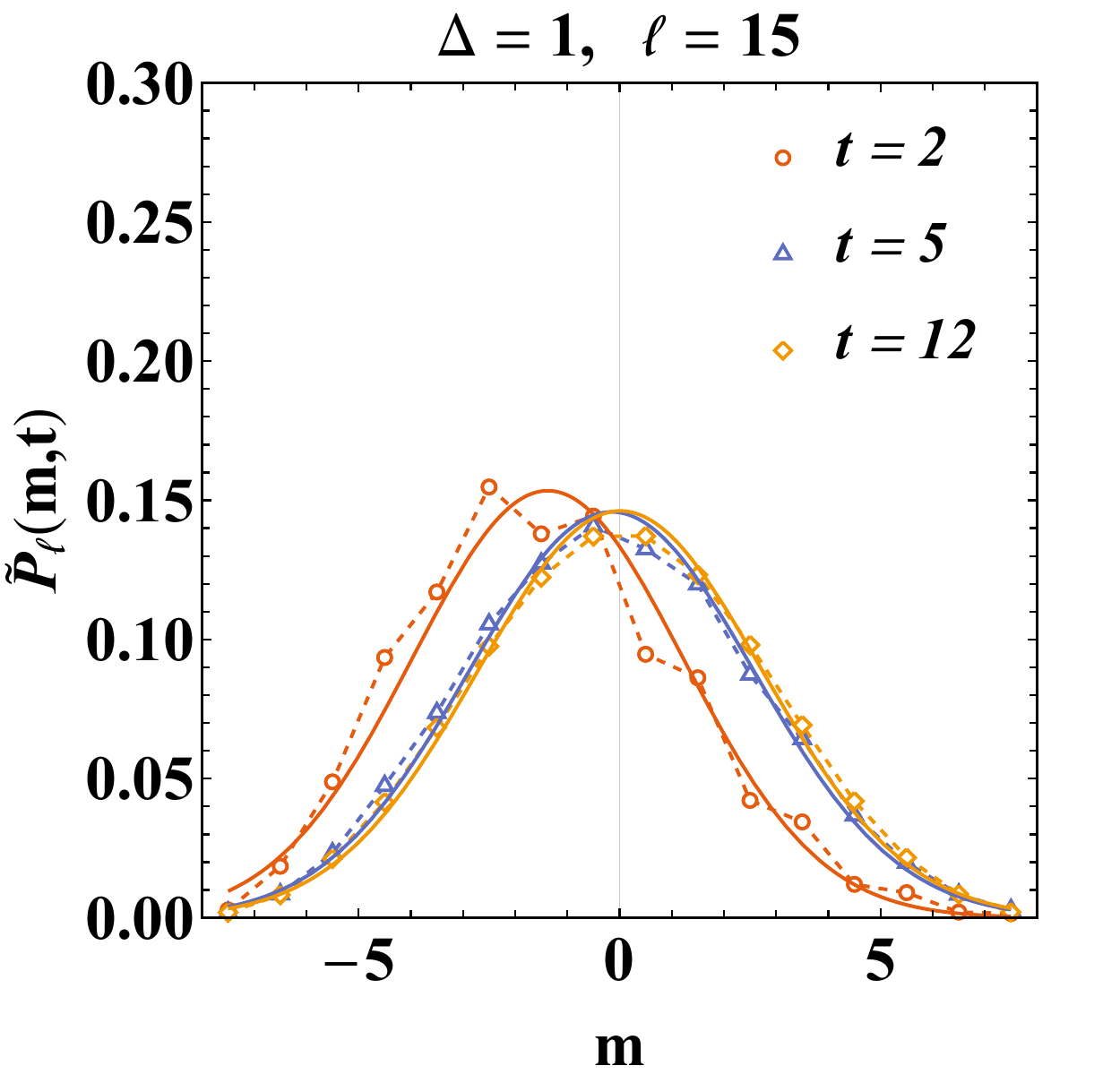}\\
\includegraphics[width=0.2325\textwidth]{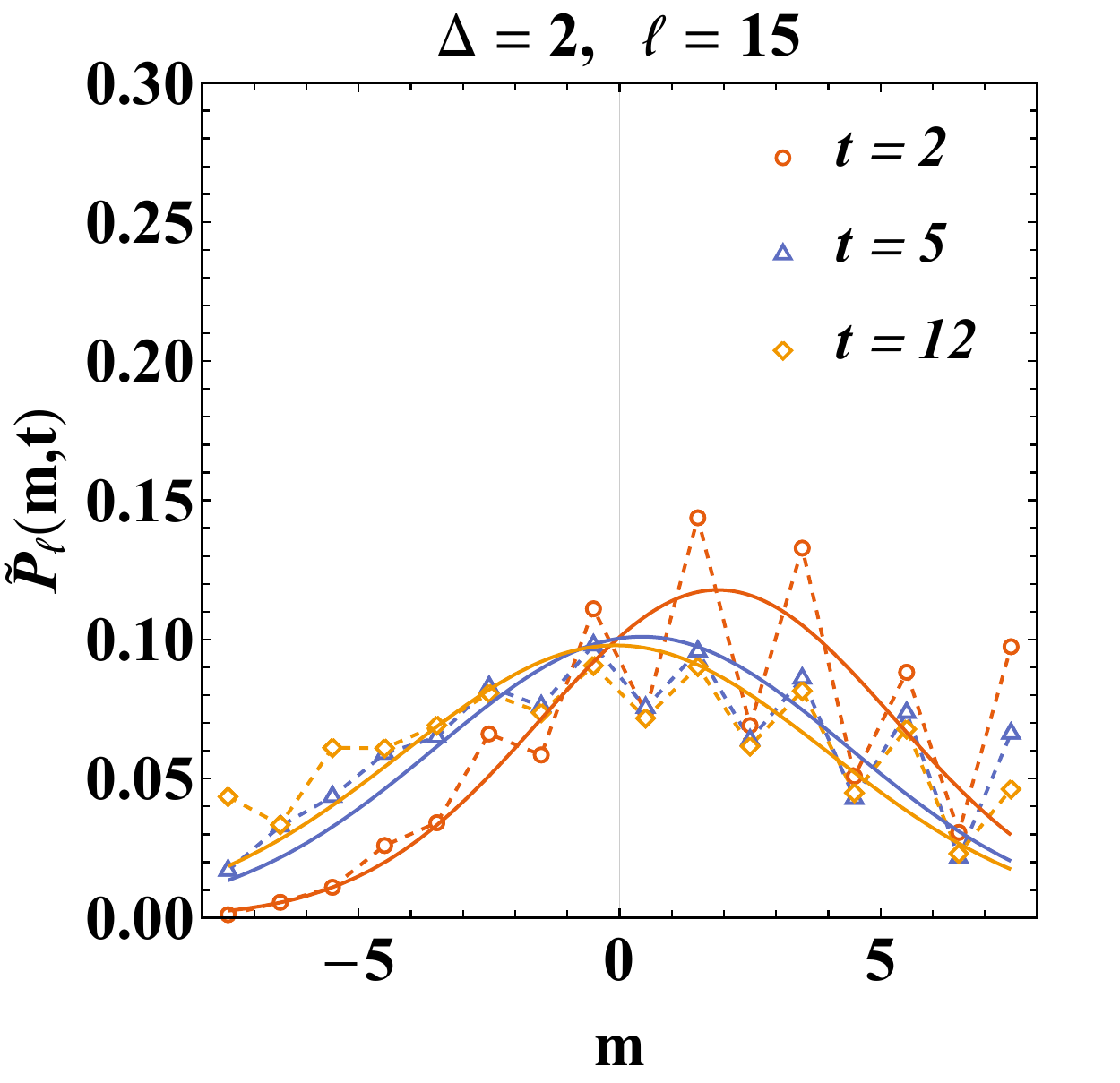}
\includegraphics[width=0.2325\textwidth]{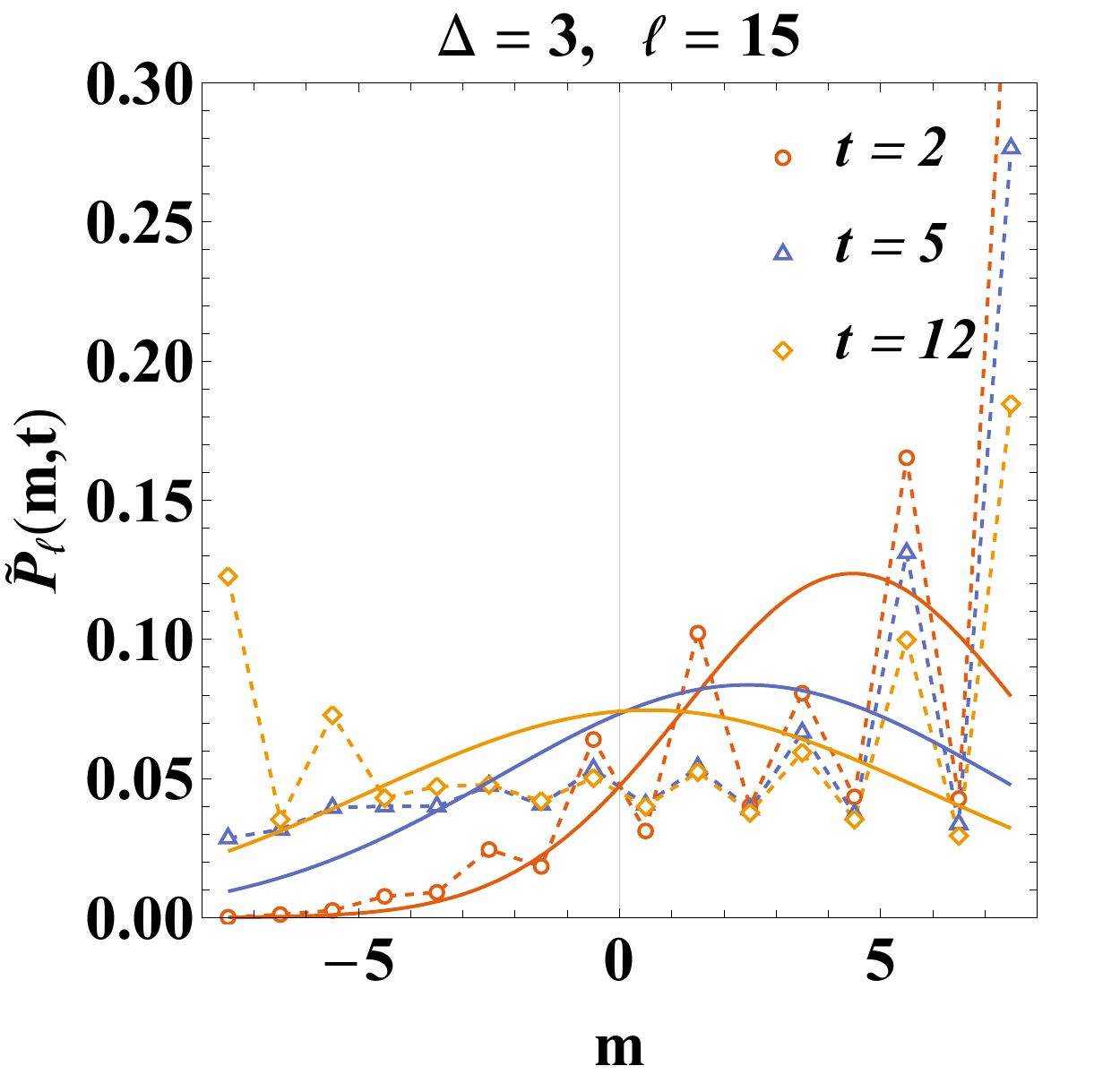}
\caption{\label{fig:rP_snapshot}
Snapshots of the rescaled PDF
in Fig.~\ref{fig:P_density} at fixed times $t=5,\, 10,\, 15$. 
The numerical data (symbols/dashed lines) are compared to the Gaussian approximation
(\ref{eq:P_gauss}) (full lines).
}
\end{center}
\end{figure}
We observe that the probability distribution depends strongly on
$\Delta$: for small values of $\Delta$ the antiferromagnetic
short-ranged order melts quickly and $P_\ell(m,t)$ is narrowly
peaked around its average, which exhibits a damped oscillatory
behaviour around zero~\cite{BPGDA09}. The behaviour for
$\Delta\agt 2$ is very different: short-ranged order persists for some
time while the probability distribution broadens and becomes more
symmetric in $m$. This nicely chimes with the expectation (see below)
that in the stationary state reached at late times the probability
distribution to become symmetric in $m$. 
In Fig.~\ref{fig:rP_snapshot} we plot the weights of $P_\ell(m,t)$ at
several times and compare them to a Gaussian approximation based on
the first two moments $\bar{m}(t)=\langle\Phi_t|M_\ell|\Psi_t\rangle$,
$\sigma^2(t)= \langle\Phi_t|M_\ell^2|\Psi_t\rangle-\bar{m}^2(t)$
\be
\label{eq:P_gauss}
P_{\ell}(\mu, t)=
\frac{1}{\sqrt{2\pi\sigma(t)}}  
\exp \left\{ - \frac{[\mu - \bar{m}(t)]^2}{2 \sigma(t)} \right\} .
\ee
We see that at $\Delta=0$ the probability distribution is
approximately Gaussian at all times, while for $\Delta=2,3$ it
exhibits a pronounced even/odd structure at short times and even at
the latest times shown is strongly non-Gaussian.
\paragraph{``Small''- $\Delta$ regime. ---}\label{sec:short}
At small values of $\Delta$ and short and intermediate times we can use a
time-dependent self-consistent mean-field approximation to determine
the evolution of $P_\ell(m,t)$. We first map the Hamiltonian \fr{eq:H}
to a model of spinless fermions by means of a Jordan-Wigner
transformation, where we use the positive (negative) z-direction in
spin space as quantization axis for even (odd) sites (see
Supplementary material). This results in a spinless fermion
Hamiltonian  
\be
H_\Delta=\sum_j \frac{1}{2}\left[c^\dagger_jc^\dagger_{j+1}+{\rm h.c.}\right]
+\Delta \, n_j(1-n_{j+1}),
\ee
where $n_j=c^\dagger_jc_j$ and
$\{c_j, c^\dagger_{k}\}=\delta_{j,k}$. The staggered subsystem
magnetization maps to 
$M_\ell= \sum_{j=1}^{\ell} ( 1/2-n_j)$, while the initial N\'eel state
maps to the fermion vacuum $|\Psi_0\rangle=|0\rangle$. Our
self-consistent approximation corresponds to the replacement  
\bea
n_jn_{j+1}&\rightarrow& 
\left[\langle c^\dagger_jc^\dagger_{j+1}\rangle_tc_{j+1}c_j
-\langle c^\dagger_{j+1}c_{j}\rangle_tc^\dagger_{j}c_{j+1}
+{\rm h.c.}\right]\nn
&+&\langle n_j\rangle_tn_{j+1}+
\langle n_{j+1}\rangle_tn_j,
\label{decouple}
\eea
which leads to an explicitly time-dependent Hamiltonian $H_{\rm MF}(t)$,
cf. Ref.~\onlinecite{SotiriadisCardy}. The expectation values in \fr{decouple}
are calculated self-consistently
$\langle.\rangle_t=\langle\Psi_t|.|\Psi_t\rangle$, where
\be
|\Psi_t\rangle=
T\exp\left[-i\int_0^tdt'H_{\rm MF}(t')\right]
|0\rangle.
\ee
Following Ref.~\onlinecite{Groha} we can express the characteristic
function of $P_\ell(m,t)$ as a determinant of a $2\ell\times2\ell$
matrix (see Supplementary material), which is easily evaluated
numerically. This provides us with exact results at $\Delta=0$ for all
times, \emph{cf.} Fig.~\ref{fig:short0}, and a highly accurate
short-time approximation even for $\Delta=3$ as is shown in 
Fig.~\ref{fig:short3}.

\begin{figure}[t!]
\begin{center}
\includegraphics[width=0.225\textwidth]{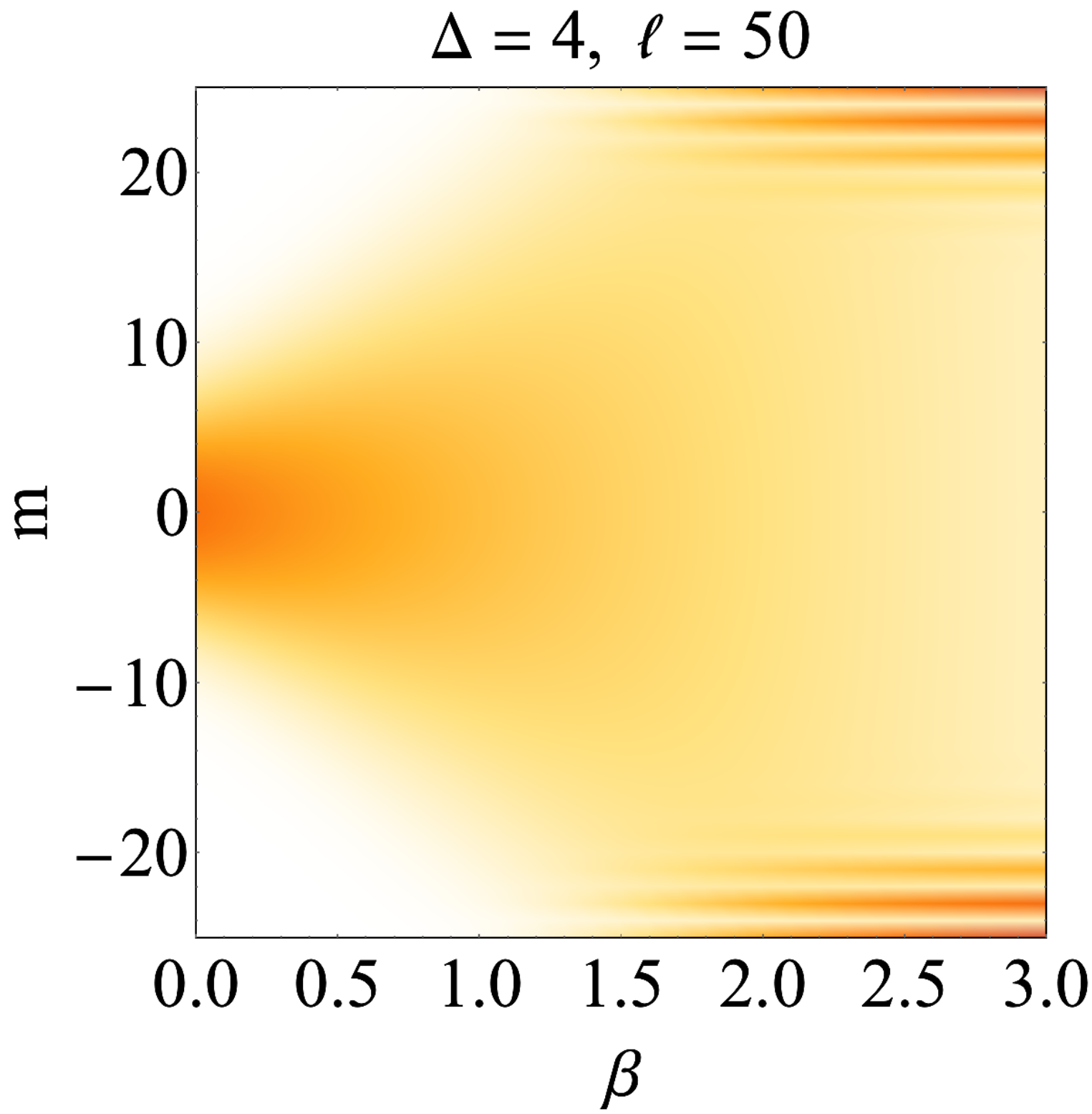}
\includegraphics[width=0.225\textwidth]{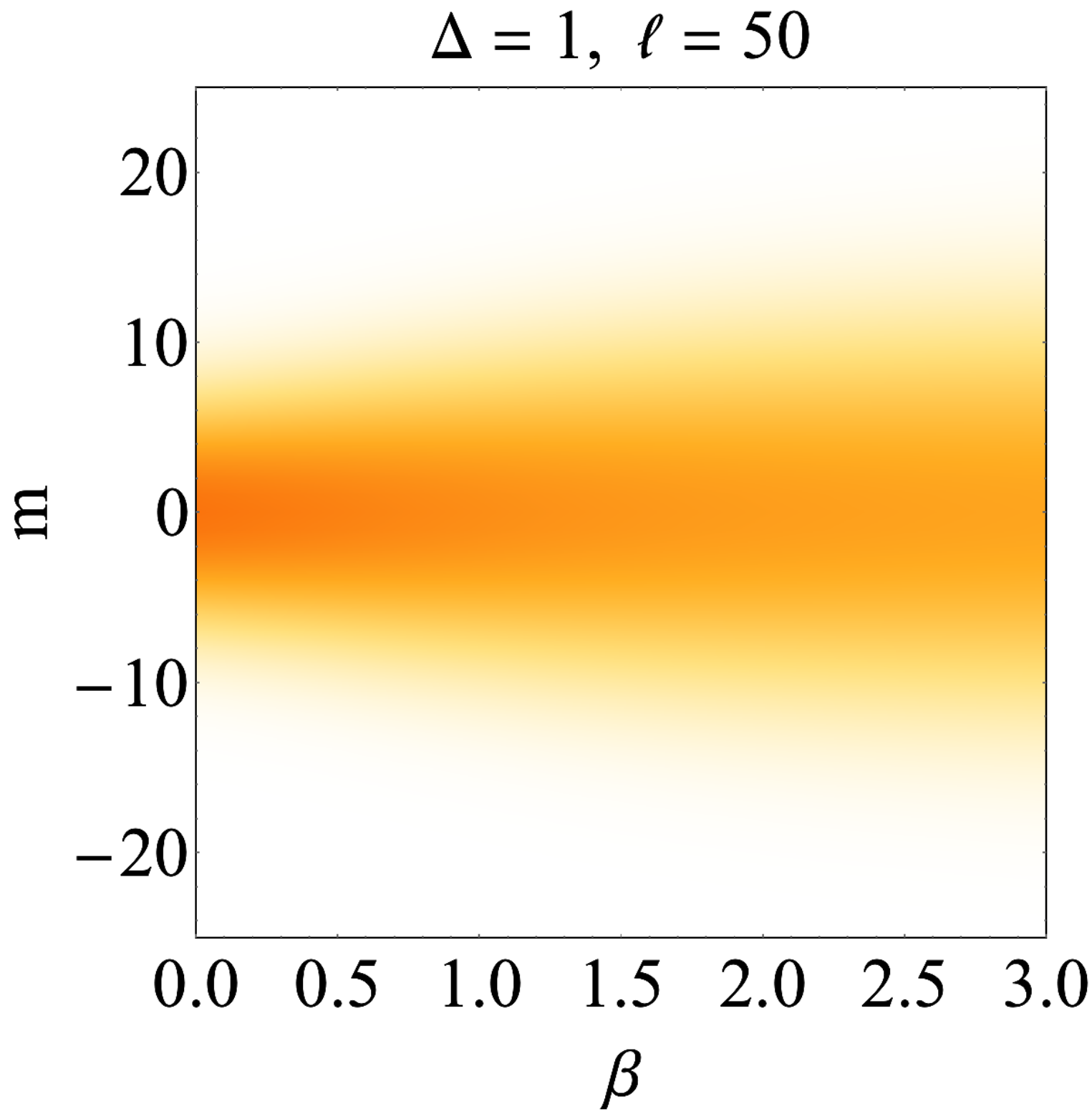}
\raisebox{16pt}{\includegraphics[height=0.18\textwidth]{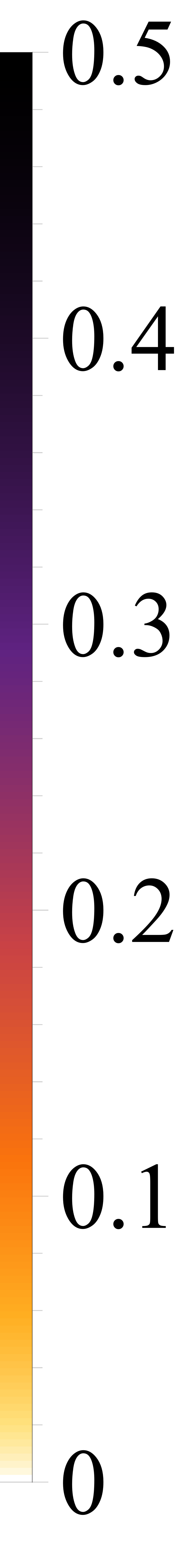}}
\caption{\label{fig:finiteT}
Density plot of the PDF for the XXZ chain at finite temperature $1/\beta$ for subsystem
size $\ell=50$ and $\Delta=4$ (left panel); $\Delta=1$ (right panel).
}
\end{center}
\end{figure}

\paragraph{Late times and large $\Delta$. ---}\label{sec:stat}
We now turn to the behaviour at late times after the quench.
The stationary state is characterized by a finite correlation length
$\xi(\Delta)$. On length scales $\ell\alt\xi(\Delta)$ we expect
short-ranged antiferromagnetic order to remain, while it will have
melted at scales $\ell>\xi(\Delta)$. We also expect the
spin-rotational symmetry by 180 degrees around the x-axis to be
restored in the stationary state as we are dealing with a one
dimensional system with short-range interactions. The situation is
completely analogous to that at finite temperatures -- in fact 
adding a very small integrability-breaking term to the Hamiltonian would
result in a steady state that is very close to the thermal state of
the XXZ chain~\cite{QANeel}. In contrast to the steady state after our
quench, the probability distribution of the staggered subsystem
magentization at finite temperature $P_\ell(m,\beta)$ can be computed
by matrix product state methods and for the aforementioned reasons it
is instructive to consider it. Results for two values of $\Delta$ are
shown in Fig.~\ref{fig:finiteT}. 
We see that the probability distributions are symmetric in $m$,
reflecting the unbroken symmetry of rotations around the x-axis by 180
degrees. At $\Delta=4$ we further observe that when the subsystem size
exceeds the thermal correlation length $\xi_\Delta(\beta)$
antiferromagnetic short-ranged order melts and we obtain a Gaussian
probability distribution centred around $m=0$. On the other 
hand, for $\ell\alt\xi_\Delta(\beta)$ the probability distribution is
very broad and peaked at the maximal values $\pm\ell/2$, signalling the presence
of both kinds of antiferromagnetic short-ranged order. For $\Delta=1$
the thermal correlation length is smaller than one lattice site in the
temperature regime shown, which is why no traces of short-range order
are visible and the probability distribution is a Gaussian centred
around $m=0$.

The large-$\Delta$ regime is characterized by a low density of
excitations and it is therefore possible to understand the behaviour
observed above by combining a $1/\Delta$-expansion with a
linked-cluster expansion along the lines of
Refs \onlinecite{EK:finiteT,JEK,GKE,PT10,ST12,CEF1,CEF2,SE:Ising,sine-Gordon}. As 
the physics we wish to describe is not tied to integrability, and the
non-integrable case is easier to discuss, we focus on the latter. We
consider the regime $\Delta\gg 1$ and break integrability by adding
a small perturbation to the Heisenberg Hamiltonian, e.g. consider time
evolution under $H=H_\Delta+\Delta^{-n}V$, where $n$ is a positive
integer and $V$ some perturbation involving short-ranged spin-spin
interactions that has the same symmetries as $H_\Delta$. We define
linked clusters following the general formalism of
Ref. \onlinecite{EK:finiteT} and then implement a $1/\Delta$-expansion
through a unitary transformation
$\widetilde{H}=e^{iS}He^{-iS}$~\cite{SchriefferWolf}, see
the Supplementary Material for details. The result is an expansion of
the stationary state density matrix of the form 
\be
\rho_{\rm SS}=\sum_{j\geq 0}\rho_{\rm SS}^{(j)}\ ,\
\rho_{\rm SS}^{(j)}={\cal O}\big(e^{-\beta_{\rm eff} j\Delta/2}\big)\ ,
\label{rhoSS}
\ee
where $\rho_{\rm SS}^{(j)}$ are given as power series in
$1/\Delta$. The leading term in the expansion is $\rho_{\rm
SS}^{(0)}=\frac{1}{2}\sum_{\sigma=\pm}|{\rm
GS}_\Delta^\sigma\rangle\langle {\rm GS}_\Delta^\sigma|$, where
$|{\rm GS}_\Delta^\sigma\rangle$ are the two ground states of the
model at anisotropy $\Delta$. The small parameter $e^{-\beta_{\rm eff}
\Delta/2}$ is proportional to the density of domain-wall excitations
over the ground states at large $\Delta$. The expansion \fr{rhoSS} of
the steady-state density matrix leads to a corresponding expansion of
the probability distribution of the staggered subsystem magnetization
$P_\ell(m,\infty)=\sum_jP^{(j)}_{\ell}(m)$ (see Supplemental Material)
\bea
P^{(0)}_{2\ell}(m)&=&\delta(\ell-|m|)\left[\frac{1}{2}-\frac{2\ell+1}{8\Delta^2}\right]
+\delta(\ell-1-|m|)\frac{1}{4\Delta^2}\nn
&+&\delta(\ell-2-|m|)\frac{2\ell-1}{8\Delta^2}+o(\Delta^{-2}),
\eea
\bea
P^{(1)}_{2\ell}(m)&=&
e^{-\frac{\beta_{\rm eff}\Delta}{2}}I_0(\beta_{\rm eff})
\bigg[\frac{1-\ell}{2}\delta(\ell-|m|)\nn
&+&\sum_{j=1}^\ell \delta(\ell-j-|m|)\bigg]
+\dots,
\label{Pinfty}
\eea
where the dots denote subleading terms in $1/\Delta$. The
expansions \fr{Pinfty} hold as long as the subsystem size $2\ell$ is
small compared to the correlation length in $\rho_{\rm SS}$ and establish
that for large anisotropies $\Delta$ the probability distribution in
the steady state is symmetric in $m$ and close to the average over the
two ground states. In addition there is an exponentially suppressed
''background'' contribution arising from a dilute gas of domain walls.

\begin{figure}[t!]
\begin{center}
\includegraphics[width=0.4\textwidth]{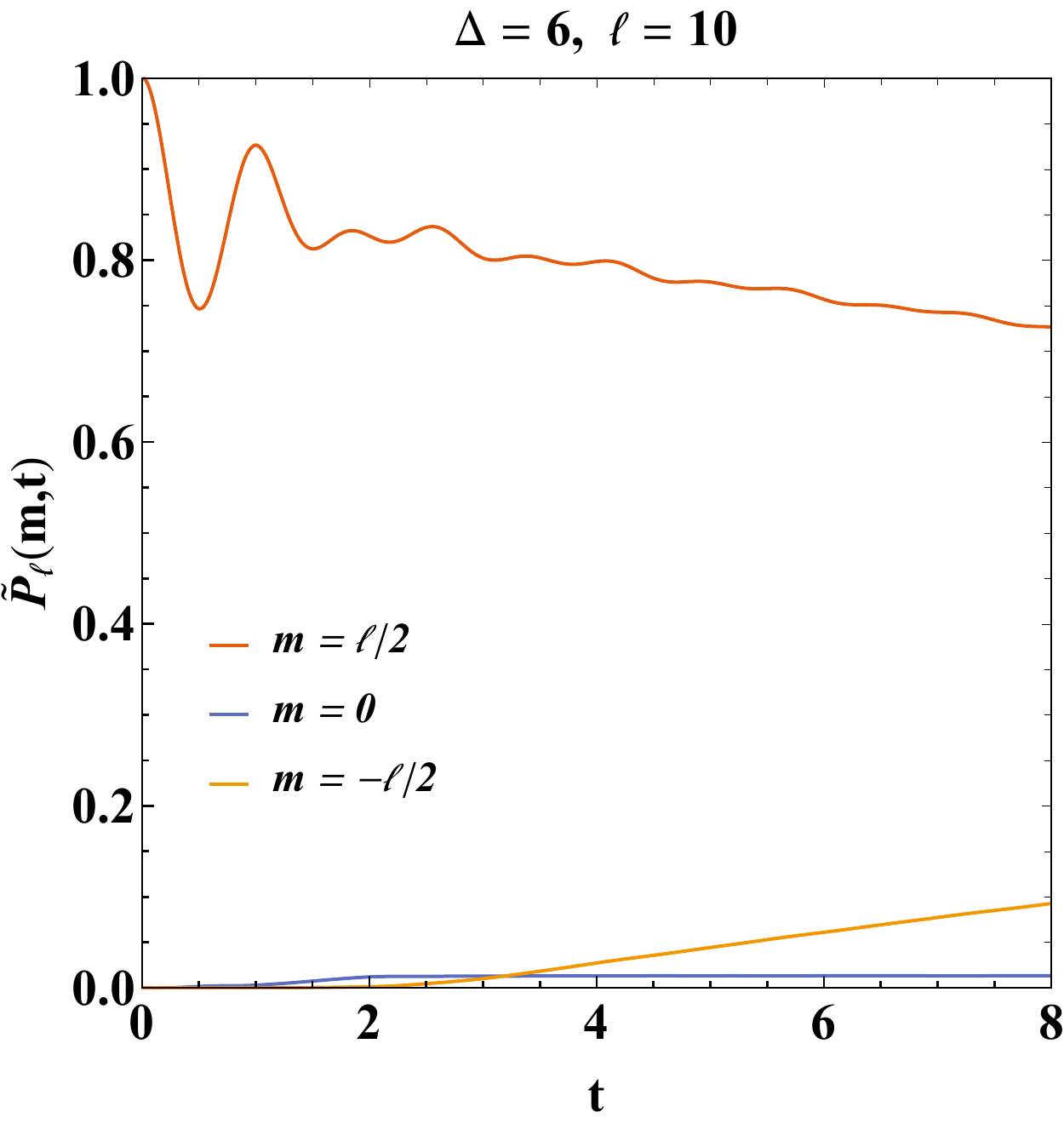}
\caption{Weights $\tilde{P}_\ell(m,t)$ for $\ell=10$ and a quench from
the classical N\'eel state to a Heisenberg chain with $\Delta=6$. We
observe a linear growth (decrease) in time of the weights for
$m=-\ell/2$ ($m=\ell/2$), indicating that the symmetrization of the PDF
is driven by ballistic propagation of quasi-particles.
}
\label{fig:symm}
\end{center}
\end{figure}

\paragraph{``Symmetrization'' of the PDF in time. ---}
\label{sec:restore}
A characteristic feature of the time-evolution of $P_\ell(m,t)$ is
that it becomes increasingly symmetric in $m$. In order to ascertain
the associated time scale in the most interesting large-$\Delta$
regime it is useful to compare the probabilities for $M_\ell$ to be
maximal ($\ell/2$) or minimal ($-\ell/2$) respectively. Results for
$\Delta=6$ are shown in Fig.~\ref{fig:symm}.
We see that $P_\ell(-\ell/2,t)$ grows linearly in time, while
$P_\ell(\ell/2,t)$ shows a corresponding linear decrease. For the
integrable XXZ chain the associated velocity is expected to be the
maximal group velocity of elementary excitations over the stationary 
state~\cite{BEL}. In presence of weak integrability-breaking interactions in
the large-$\Delta$ regime we expect qualitatively similar prethermal
behaviour~\cite{prethermal}.

\paragraph{Conclusions. ---}
\label{sec:conclusion}
We have considered the full quantum mechanical probability
distribution of the staggered subsystem magnetization $P_\ell(m,t)$
after a quantum quench from a classical N\'eel state to the spin-1/2
Heisenberg XXZ chain. We have shown that $P_\ell(m,t)$ provides
detailed information how short-range antiferromagnetic order melts and
have shown how to understand our numerical findings by analytical
approaches valid in certain limits. Our setup should be realizable in
cold-atom experiments like the ones by the Harvard group~\cite{Greiner}.

\paragraph{Acknowledgments. ---}\label{sec:acknowledgments}
We are grateful to P. Calabrese, A. De Luca and M. Fagotti for
stimulating discussions and to the Erwin Schr\"odinger International
Institute for Mathematics and Physics for hospitality and support
during the programme on \emph{Quantum Paths}. M.C. thanks
the Galileo Galilei Institute in Florence for hospitality
during the workshop ``Entanglement in quantum systems''.
This work was supported by the the European Union's Horizon 2020 under
the Marie Sklodowska-Curie grant agreement No.~701221 NET4IQ (M.C and
F.H.L.E.), the EPSRC under grant EP/N01930X (F.H.L.E.), the National Science
Foundation under Grant No. NSF PHY-1748958 (F.H.L.E.) and by the BMBF and
EU-Quantera via QTFLAG (M.C.).


\appendix*
\section{SUPPLEMENTARY MATERIAL}
\setcounter{equation}{0}
\subsection{Self-consistent time dependent mean-field approximation}
In the small-$\Delta$ regime the quench dynamics of the Heisenberg XXZ
chain can be analyzed by means of a self-consistent time-dependent
mean-field theory~\cite{SotiriadisCardy} as we will now demonstrate. For convenience we
rotate the spin-quantization axis on all odd sites, which maps the
classical N\'eel state on the saturated ferromagnetic state with all
spins up. This transformation is induced by $P = \prod_{j\;{\rm
odd}} \sigma^{x}_{j}$ and we have
\bea
|\Psi_0\rangle & \equiv & P|\Phi_{0}\rangle = |\uparrow\uparrow\dots\uparrow\rangle, \nn
S_{\ell} &\equiv& P M_{\ell} P = \sum_{j=1}^{\ell} S^{z}_{j}, 
\eea
while the Hamiltonian maps onto
\bea
H'_{\Delta} &\equiv& P H_{\Delta}P\nn
& =& \sum_{j} S^{x}_{j}S^{x}_{j+1} - S^{y}_{j}S^{y}_{j+1} - \Delta
S^{z}_{j}S^{z}_{j+1}\ .
\eea
The probability distribution of the staggered subsystem magnetization
in the XXZ chain is then expressed as
\be
P_\ell(m,t)=\langle \Psi_t | 
\delta( S_{\ell} - m)
| \Psi_t \rangle\ ,\quad
|\Psi_t\rangle=e^{-iH'_{\Delta}t}|\Psi_0\rangle\ .
\ee
Performing the Jordan-Wigner transformation to spinless fermions
$\{c_{i},c^{\dag}_{j}\}=\delta_{ij}$
\be
S^{+}_{j} = \prod_{i}^{j-1}(1-2n_{i})c_{i}, \quad
S^{z}_{j} = \frac{1}{2} - n_{j},  
\ee
the Hamiltonian becomes (up an unimportant constant)
\be\label{eq:tildeH_fermi}
H'_{\Delta} = 
\sum_{j}
\frac{1}{2} \left[ c^{\dag}_{j}c^{\dag}_{j+1} - c_{j} c_{j+1} \right]
+\Delta \, n_j\left[1-n_{j+1}\right],
\ee
where $n_{i}= c^{\dag}_{i}c_{i}$. The initial state maps onto the
fermion vacuum, i.e.
\be
|\Psi_0\rangle=|0\rangle.
\ee
We now decouple the four-fermion
interaction in a self-consistent fashion
\bea
c^{\dag}_{j}c^{\dag}_{j+1}c_{j+1}c_{j} & \approx &
\langle n_j \rangle_{t} \, n_{j+1}
+\langle n_{j+1} \rangle_{t} \, n_j\nn
&-&\langle c^{\dag}_{j} c_{j+1} \rangle_{t} \, c^{\dag}_{j+1} c_{j}
-\langle c^{\dag}_{j+1} c_{j} \rangle_{t} \, c^{\dag}_{j} c_{j+1} \nn
& + & \langle c^{\dag}_{j} c^{\dag}_{j+1} \rangle_{t} \, c_{j+1} c_{j}
+\langle c_{j+1} c_{j} \rangle_{t} \, c^{\dag}_{j} c^{\dag}_{j+1},\nn
\label{SCMF}
\eea
which leads to a time-dependent mean-field Hamiltonian
\bea
H_{\rm
MF}(t)&=&\sum_{j}[A_j(t)c_{j+1}c_j+B_j(t)c^\dagger_{j+1}c_j+{\rm h.c.}]\nn
&+&\sum_jh_j(t)n_j\ .
\eea
Here we have defined
\bea
A_j(t)&=&\frac{1}{2}-\Delta\langle
c^\dagger_jc^\dagger_{j+1}\rangle_t\ ,\nn
B_j(t)&=&\Delta\langle c^\dagger_jc_{j+1}\rangle_t\ ,\nn
h_j(t)&=&\Delta\left[1-\langle n_{j-1}\rangle_t-\langle n_{j+1}\rangle_t\right].
\label{SCC}
\eea
The expectation values at time $t$ in (\ref{SCMF}) and (\ref{SCC}) are
calculated self-consistently
\bea
\langle {\cal O}\rangle_t&=&\langle 0|U^\dagger(t){\cal
O}U(t)|0\rangle\ ,\nn
U(t)&=&T\exp\left[-i\int_0^tdt'H_{\rm MF}(t')\right].
\eea
The self-consistent mean-field approximation obtained in this way is
expected to work well for small values of $\Delta$ and sufficiently
short times. As we will see it works quite well even for intermediate
values of $\Delta$.

The Heisenberg equations of motion in our self-consistent mean-field
approximation are
\bea
-i\frac{d}{dt} c_{j}(t) &=& U^\dagger(t)[ H'_{\Delta},c_{j}]U(t)\nn
&=&A_{j-1}^*(t) c^{\dag}_{j-1}(t)- A_j^*(t) c^{\dag}_{j+1}(t)\nn
& - &B_{j-1}(t)  c_{j-1}(t) -B_j^*(t) c_{j+1}(t)\nn
&  -&  h_j(t) c_{j}(t).
\eea
It is convenient to cast them in matrix form
\be
\frac{d}{dt} {\bf d}(t) = i \, \mathbb{T}_{\Delta}[\mathbb{C}(t)] \cdot {\bf d}(t),
\ee
where ${\bf d}^{\dag} \equiv (c^{\dag}_{1}, \dots,
c^{\dag}_{L},c_{1},\dots, c_{L})$ and we have defined a $2L\times 2L$ correlation
matrix $\mathbb{C}(t)$ by
\be
\left[\mathbb{C}(t)\right]_{nm}=\langle 0|d_n(t) d^{\dag}_m(t)|0\rangle.
\ee
The time evolution of the correlation matrix is governed by a
first-order differential equation 
\be
\frac{d\mathbb{C}(t)}{dt} =  i \mathbb{T}_{\Delta}[\mathbb{C}(t)] \cdot \mathbb{C}(t)
 -i \mathbb{C}(t) \cdot \mathbb{T}^{\dag}_{\Delta}[\mathbb{C}(t)].
\label{corrmat}
\ee
Using that our initial state is the fermion vacuum we have
\be
\left[\mathbb{C}(0)\right]_{nm}=\begin{cases}
\delta_{n,m} &\text{if } n>L\\
0 &\text{else}.
\end{cases}
\ee
The correlation matrix at time $t$ can now be straightforwardly obtained by
numerically integrating \fr{corrmat}.
By comparing the results to iTEBD computations we observe very good
agreement for small values of $\Delta$. For large values $\Delta > 1$
the approximation still gives a fair description of the dynamics at
short times $t\alt 1$.
\subsubsection{Full counting statistics}
A nice feature of the self-consistent mean-field approximation
described above is that it allows us to determine the
probability distribution $P_\ell(m,t)$. To that end we define the
associated characteristic function by
\bea
P_\ell(m,t)&=&\int_{-\infty}^\infty\frac{d\theta}{2\pi}e^{-im\theta}\
F_\ell(\theta,t)\ ,\nn
F_\ell(\theta,t)&=&\langle 0|U^\dagger(t)
e^{i\theta S_\ell}U(t)|0\rangle\ .
\eea
Following Ref.~\onlinecite{Groha} we can derive a determinant
representation for $F_\ell(\theta,t)$ that can be efficiently
evaluated. We start by introducing Majorana fermion operators by
\be
a^{x}_{j} \equiv c^{\dag}_{j} + c_{j},\quad a^{y}_{j} \equiv i(c^{\dag}_{j}-c_{j}).
\ee 
These have anticommutation relations $\{a^{\alpha}_{i},a^{\beta}_{j}\}
= 2 \delta_{\alpha\beta}\delta_{ij}$ and the observable of interest is
bilinear in them 
\be
S_\ell=\sum_{j=1}^\ell \frac{i}{2} \, a^{y}_{j} a^{x}_{j}.
\ee 
Next we introduce an auxiliary reduced density matrix acting on sites
$1$,\dots,$\ell$ by
\be
\rho^{\rm aux}_\ell=\frac{1}{Z}e^{i\theta S_\ell}=\frac{1}{Z} \exp\left[\frac{\theta}{4}{\bf
a}^{\dag} \mathbb{W} {\bf a}\right], 
\ee
where ${\bf a}^{\dag} = (a^{x}_{1},\dots,a^{x}_{\ell},a^{y}_{1},\dots,
a^{y}_{\ell})$, $\mathbb{W} = i \sigma^{y}\otimes \mathbb{1}_{\ell}$ and
\be
Z = {\rm Tr} \left[ {\rm
e}^{i\theta \sum_{j=1}^{\ell}S^{z}_{j}} \right] =
[2\cos(\theta/2)]^{\ell}.
\ee
The characteristic function can then be expressed as
\be\label{eq:F_meanfield}
F_{\ell}(\theta,t) = Z \; {\rm Tr} \big[ 
\rho_\ell^{\rm aux}
\rho_{\ell}(t)\big],
\ee
where $\rho_\ell(t)$ is the reduced density matrix of the subsystem
$[1,\ell]$ at time $t$
\be
\rho_\ell(t)={\rm Tr}_{[\ell+1,L]}\left[U(t)|0\rangle\langle 0|U^\dagger(t)\right].
\ee
Finally we use an identity for the trace of the product of two Gaussian density
matrices proved in Ref.~\onlinecite{Fagotti}
\be
{\rm Tr}\left[\rho_1\rho_2\right]=\sqrt{{\rm
det}\left(\frac{1+\Gamma_1\Gamma_2}{2}\right)}\ ,
\ee
where $\Gamma_j$ are correlation matrices defined by
\be
\left(\Gamma_j\right)_{nm}={\rm Tr}\left[\rho_j\ a_ma_n\right]-\delta_{n,m}.
\ee
In our case the two correlation matrices are
\bea
\Gamma^{\rm aux}_\ell &=& \tanh\left(\frac{\theta\mathbb{W}}{2}\right)
=\tan\left(\frac{\theta}{2}\right)\mathbb{W}\ ,\nn
\Gamma_\ell(t) &=&\langle 0|{\bf a}(t){\bf
a}^\dagger(t)|0\rangle-\mathbb{1}_{2\ell}\nn
&=&\mathbb{M}\mathbb{C}(t)\mathbb{M}^\dagger-\mathbb{1}_{2\ell}\ ,
\eea
where $\mathbb{C}(t)$ is the correlation matrix obtained
from \fr{corrmat} and
\be
\quad \mathbb{M}=
\begin{pmatrix}
\mathbb{1}_\ell & \mathbb{1}_\ell\\
-i\mathbb{1}_\ell & i\mathbb{1}_\ell
\end{pmatrix}.
\ee
Our final result for the characteristic function of the probability
distribution $P_\ell(m,t)$ is thus
\be
F_\ell(\theta,t)=\big[2\cos(\theta/2)\big]^\ell
\sqrt{{\rm det}\left(\frac{1+\Gamma^{\rm aux}_\ell\Gamma_\ell(t)}{2}\right)}\ .
\label{charfn}
\ee
Combining (\ref{charfn}) and (\ref{corrmat}) we can obtain numerically
exact results in the framework of our time-dependent self-consistent
mean-field approximation for $F_\ell(\theta,t)$ in the thermodynamic
limit and large subsystem sizes.
\section{Combined linked-cluster and $1/\Delta$ expansions}
In the large-$\Delta$ regime we can gain insights about the late time
behaviour after the quench by means of combined linked-cluster and
$1/\Delta$ expansions. We again carry out a rotation of the
spin-quantization axis, which leads us to a Hamiltonian of the form
\be
H'_{\Delta}=-J\Delta\sum_{j}\tau^z_j\tau^z_{j+1}+2J\sum_{j}\tau^+_j\tau^+_{j+1}+
\tau^-_j\tau^-_{j+1}\ .
\ee
Here $\tau^\alpha_j$ are Pauli matrices and we have introduced an
energy scale $J$, which should be set to $1/4$ in order to arrive
at the conventions used in the main text. In the
rotated basis the staggered subsystem magnetization maps onto
\be
S^z_\ell=\frac{1}{2}\sum_{j=1}^\ell \tau^z_j\ ,
\ee
and our goal is to calculate
\be
P_\ell(m,\infty)=\lim_{t\to\infty}
\int_{-\infty}^\infty \frac{d\theta}{2\pi}\ e^{-im\theta}
\langle\uparrow|e^{i\theta S^z_\ell(t)}|\uparrow\rangle,
\ee
where $|\uparrow\rangle$ is the saturated ferromagnetic state. As the
physics we are interested in does not rely on the integrability of the
spin-1/2 XXZ chain and the non-integrable case is easier to analyze,
we proceed by adding a short-ranged integrability-breaking interaction
$V$ that does not break any of the discrete symmetries of
$H'(\Delta)$ and consider 
\be
H=H'_{\Delta}+\Delta^{-n}V\ ,
\label{Htot}
\ee
where $n$ is a positive integer. In the regime $\Delta\gg 1$ we may
carry out a $1/\Delta$-expansion following
Ref.~\onlinecite{SchriefferWolf}. We consider a basis transformation
of the form 
\be
\widetilde{H}=e^{iS}He^{-iS}\ ,\quad S=\sum_{j\geq 1}S^{(j)}\Delta^{-j}\ ,
\label{S}
\ee
where $S^{(j)}$ are to be chosen in such a way that order by order in
the expansion $\widetilde{H}$ commutes with the domain-wall number
operator
\be
N_{\rm DW}=\frac{1}{2}\sum_{j=1}^L1-\tau^z_j\tau^z_{j+1}\ .
\ee
The first order term in \fr{S} is
\be
S^{(1)}=\sum_k\frac{1+i\tau^z_{k-1}}{2}\left[\tau^+_k\tau^+_{k+1}-\tau^-_k\tau^-_{k+1}\right]\frac{1+i\tau^z_{k+2}}{2}\ .
\ee
In the transformed basis the Hamiltonian has the following expansion
\bea
\widetilde{H}&=&-J\Delta\sum_j\tau^z_j\tau^z_{j+1}+J\sum_j\tau^+_j\tau^+_{j+1}+
\tau^-_j\tau^-_{j+1}\nn
&-&J\sum_j\tau^z_{j-1}\left[\tau^+_j\tau^+_{j+1}+\tau^-_j\tau^-_{j+1}
\right]\tau^z_{j+2}+{\cal O}(\Delta^{-1}),\nn
\label{Htilde}
\eea
and domain wall number is a good quantum number. The ground states of
$\widetilde{H}$ are the saturated ferromagnetic 
states $|\uparrow\rangle=|\uparrow\uparrow\dots\rangle$ and
$|\downarrow\rangle=|\downarrow\downarrow\dots\rangle$ with energy $E_0=-J\Delta
L$ (in the original spin basis these correspond to the two N\'eel states).
Low-lying excited states involve ferromagnetic domain walls. In
order to analyze properties at finite but low energy densities by
means of the method first introduced in Ref. \onlinecite{EK:finiteT}
we need to impose open boundary conditions on \fr{Htilde} so that
excited states with an odd number of domain walls are allowed. To deal
with this complication we need to shift the subsystem for which we
determine the probability distribution to the centre of the open
chain, i.e. for even $\ell$ we take
\be
S^z_\ell\longrightarrow\frac{1}{2}\sum_{j=\frac{L-\ell}{2}+1}^{\frac{L-\ell}{2}}\tau^z_j\ .
\ee
As we are interested in $\ell$ fixed and $L\to\infty$ boundary terms
will not contribute and we will recover a translationally invariant
result. Excited states involving a single domain wall can be expanded
in a basis formed by the states
\be
|n\rangle=\bigg[\prod_{j=n+1}^L\tau^-_j\bigg]|\uparrow\rangle\ ,\quad
\ee
and their spin-reversed analogues $|\bar{n}\rangle=C|n\rangle$ where
$C=\prod_{j=1}^L\tau^x_j$. A basis of eigenstates of \fr{Htilde} in
the sector with a single domain wall is given by
$\{|Q^\pm_\alpha\rangle, C|Q^\pm_\alpha\rangle\}$, where 
\bea
|Q^+_\alpha\rangle&=&\frac{2}{\sqrt{L}}\sum_{j=1}^{L/2-1}\sin(Q_\alpha^+j)|2j\rangle\ ,\nn
|Q^-_\alpha\rangle&=&\frac{2}{\sqrt{L+2}}\sum_{j=1}^{L/2}\sin(Q_\alpha^-j)|2j-1\rangle,
\eea
and
\bea
Q^-_\alpha&=&\frac{2\pi}{L+2}\alpha\
,\quad \alpha=1,\dots,\frac{L}{2}\ ,\nn
Q^+_\alpha&=&\frac{2\pi}{L}\alpha\
,\quad \alpha=1,\dots,\frac{L}{2}-1.
\eea
The corresponding excitation energies are
\be
\epsilon(Q^\pm_\alpha)=2J\Delta+4J\cos(Q^\pm_\alpha)+{\cal O}(\Delta^{-1}).
\ee
Energy eigenstates in the original spin basis are obtained by undoing the
basis transformation \fr{S}. In particular we have (up to boundary terms)
\bea
|{\rm GS},1\rangle&=&e^{-iS}|\uparrow\rangle=C|{\rm GS},2\rangle\\
&=&|\uparrow\rangle+\frac{1}{2\Delta}
\frac{1}{\sqrt{L}}\sum_j\tau^-_j\tau^-_{j+1}|\uparrow\rangle+{\cal
O}\big(\Delta^{-2}\big).\nonumber
\label{GSs}
\eea
The density matrix
describing the steady state after a quantum quench from the classical
N\'eel state is defined by the requirement 
\be
\lim_{t\to\infty}\langle{\rm GS}_\infty^+|{\cal O}(t)|{\rm GS}_\infty^+\rangle
={\rm Tr}\left[\rho_{\rm SS}{\cal O}\right],
\ee
where ${\cal O}$ is any local operator. In our case $\rho_{\rm SS}$ is
a thermal density matrix (as we have broken integrability though the
$V$-term) and in the large-$\Delta$ regime we have (in a large, finite volume)
\cite{EK:finiteT,JEK,GKE,PT10,ST12,CEF1,CEF2,SE:Ising,sine-Gordon}
\be
\rho_{\rm SS}=\frac{1}{Z}\sum_ne^{-\beta_{\rm eff}E_n}|n\rangle\langle n|
=\sum_{j\geq 0}\rho_{\rm SS}^{(j)}
\ee
where
\be
\rho_{\rm SS}^{(j)}={\cal O}\big(e^{-\beta_{\rm eff} 2j\Delta J}\big)\ .
\ee
The first two terms of the low-density expansion involve only the
ground states and single domain wall excitations
\bea
\rho_{\rm SS}^{(0)}&=&\frac{1}{2}\sum_{a=1}^2 |{\rm GS},a\rangle\langle {\rm
GS},a|\ ,\nn
\rho^{(1)}_{\rm SS}&=&
\frac{1}{2}\sum_{\sigma=\pm}\sum_{Q_\alpha^\sigma}e^{-\beta_{\rm
eff}\ \epsilon(Q_\alpha^\sigma)}|Q^\sigma_\alpha\rangle\langle Q^\sigma_\alpha|\nn
&+&\frac{1}{2}\sum_{\sigma=\pm}\sum_{Q_\alpha^\sigma}e^{-\beta_{\rm
eff}\ \epsilon(Q_\alpha^\sigma)}C|Q^\sigma_\alpha\rangle\langle Q^\sigma_\alpha|C\nn
&&-\big[\sum_{\sigma=\pm}\sum_{Q_\alpha^\sigma}e^{-\beta_{\rm
eff}\ \epsilon(Q_\alpha^\sigma)}\big]\rho^{(0)}_{\rm SS}\ .
\label{rhoss}
\eea
By construction of the linked cluster expansion the last term
precisely subtracts the contribution that diverges in the infinite
volume limit. 
Using the explicit form \fr{GSs} of the $1/\Delta$ expansion for the
ground states we obtain an explicit expression for the leading term in
the characteristic function
\bea\label{eq:rhoSS0}
{\rm Tr}\left[\rho^{(0)}_{\rm SS}e^{i\theta
S^z_\ell}\right]&=&\cos\bigg(\frac{\ell\theta}{2}\bigg)\left[1-\frac{\ell+1}{4\Delta^2}\right]\\
&+&\cos\bigg(\frac{(\ell-2)\theta}{2}\bigg)\frac{1}{2\Delta^2}\nn
&+&\cos\bigg(\frac{(\ell-4)\theta}{2}\bigg)\frac{\ell-1}{4\Delta^2}+o(\Delta^{-2}),\nonumber
\eea
where we have used that the second order correction to the ground
state has zero overlap with the ferromagnetic states. We see that in
this contribution the probabilities of $S^z_\ell$ taking values less
than the maximal possible ones $\pm\ell/2$ are suppressed by powers of
$1/\Delta$. The subleading term in the low-density (linked cluster)
expansion is obtained from the matrix elements
\be
\langle Q_\alpha^\pm|e^{iS}e^{i\theta
S^z_\ell}e^{-iS}|Q_\alpha^\pm\rangle=
\langle Q_\alpha^\pm|e^{i\theta S^z_\ell}|Q_\alpha^\pm\rangle+{\cal O}(\Delta^{-1}).
\ee
Evaluating the matrix elements gives 
\bea\label{matrixel}
\langle Q_\alpha^+|e^{i\theta
S^z_\ell}|Q_\alpha^+\rangle&=&\frac{4}{L}\sum_{j=1}^{\frac{L}{2}-1}
\sin^2(Q_\alpha^+j)\ e^{i\theta\lambda_{2j}},\\
\langle Q_\alpha^-|e^{i\theta
S^z_\ell}|Q_\alpha^-\rangle&=&\frac{4}{L+2}\sum_{j=1}^{\frac{L}{2}}
\sin^2(Q_\alpha^-j)\ e^{i\theta\lambda_{2j-1}}, \nonumber
\eea
where for even $\ell$ we have
\be
\lambda_{j}=\begin{cases}
-\frac{\ell}{2} & \text{if }j\leq\frac{L-\ell}{2}\\
\frac{\ell}{2} & \text{if }j\geq\frac{L+\ell}{2}\\
-\frac{\ell}{2}+j-\frac{L-\ell}{2} & \text{else }
\end{cases}.
\ee
Combining \fr{matrixel} with \fr{rhoss} and turning the momentum sums
into integrals we obtain the following explicit expressions for the
first subleading contribution to the characteristic function
\bea\label{eq:rhoSS1}
{\rm Tr}\left[\rho^{(1)}_{\rm SS}e^{i\theta S^z_\ell}\right]&=&
e^{-2J\Delta\beta_{\rm eff}}I_0(4J\beta_{\rm eff})\nn
&\times&\left[\sum_{j=1}^{\ell-1}
e^{i\theta\frac{(\ell-2j)\theta}{2}}-(\ell-1)
\cos\bigg(\frac{\ell\theta}{2}\bigg)\right]\nn
&+&{\cal O}\big(e^{-2J\Delta\beta_{\rm eff}}\Delta^{-1}\big).
\eea

\begin{figure}[t!]
\begin{center}
\includegraphics[width=0.4\textwidth]{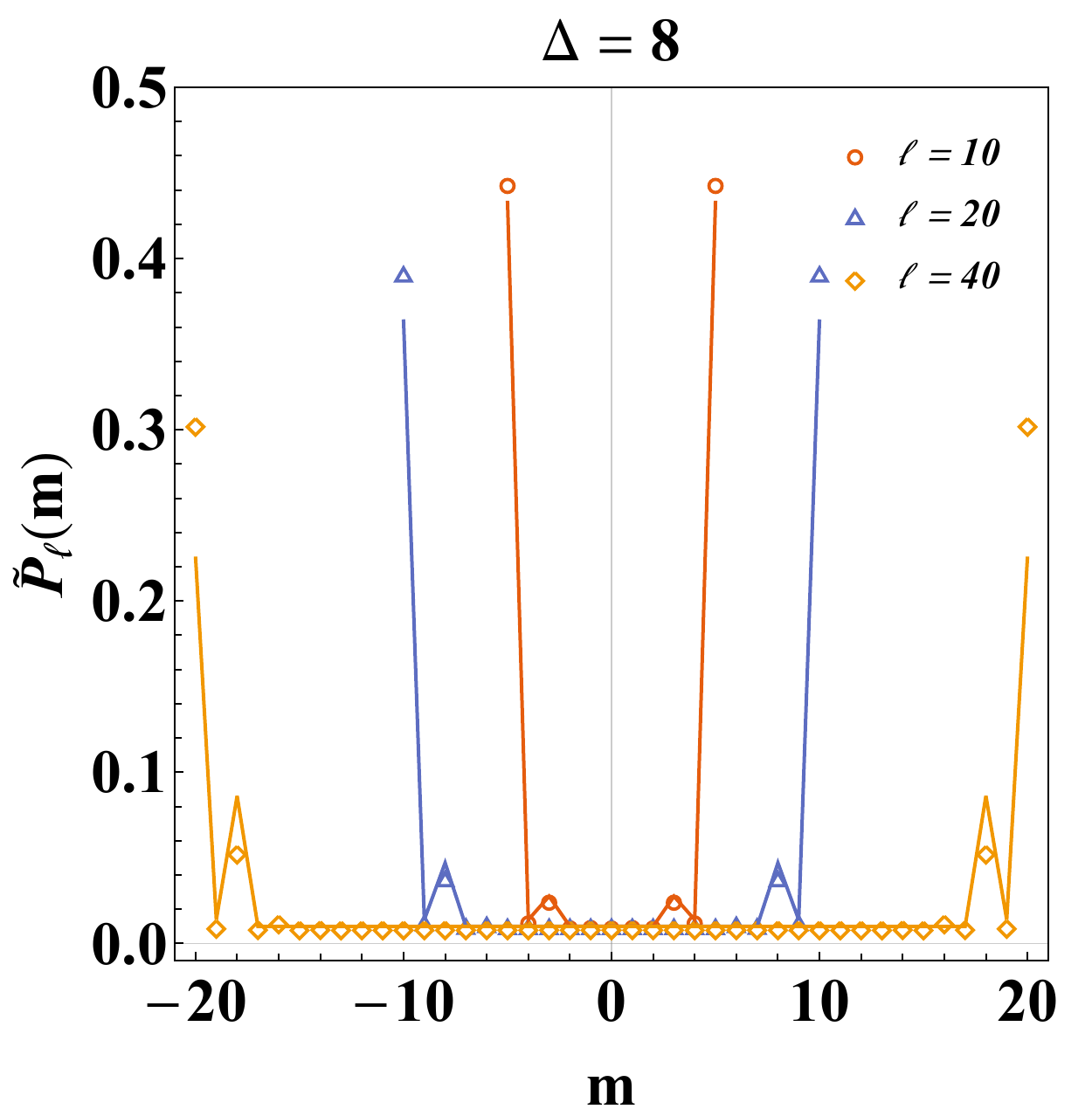}
\includegraphics[width=0.4\textwidth]{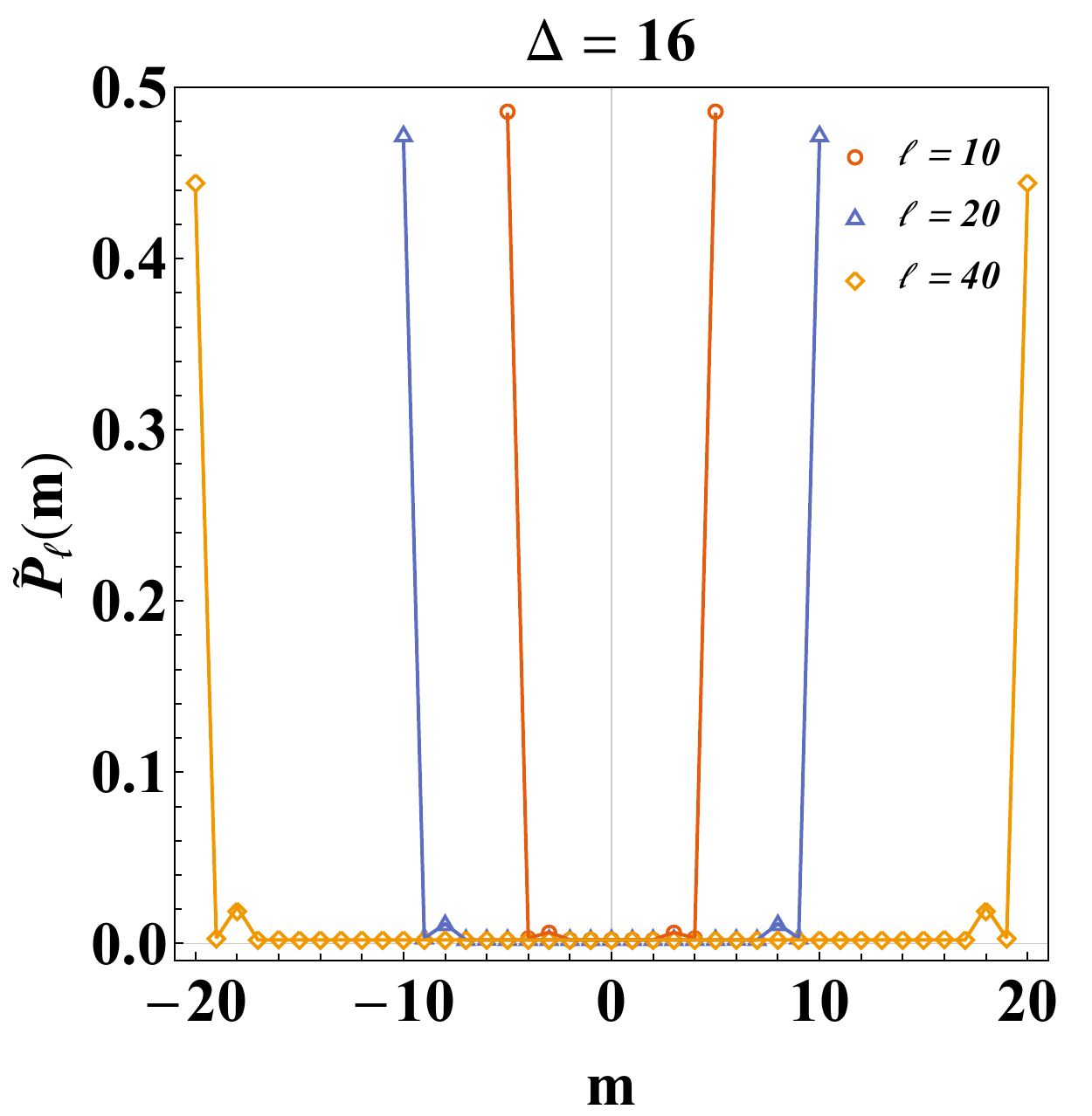}
\caption{\label{fig:largeDelta}
PDF of the staggered subsystem magnetization for several subsystem
sizes $\ell$ evaluated
in the thermal state $\rho_{\rm SS}$ with inverse effective
temperature $\beta_{\rm eff}$ fixed by the value of the energy in the
initial N\'eel state; namely $\beta_{\rm eff} = 1.241,\, 0.789$ for
$\Delta = 8,\, 16$. Symbols are exact numerical data, 
lines are the large-$\Delta$ expansion obtained by taking the Fourier transform of 
(\ref{eq:rhoSS0}) and (\ref{eq:rhoSS1}). The agreement at $\Delta=16$
is excellent.
}
\label{fig:largeDelta}
\end{center}
\end{figure}

In Fig.~\ref{fig:largeDelta} we compare the PDF in the thermal state
with inverse temperature fixed by the condition  
$
\langle{\rm GS}_{\infty}^{+}|H_{\Delta}|{\rm GS}_{\infty}^{+}\rangle
={\rm Tr}\left[\rho_{\rm SS}H_{\Delta}\right]
$ with the results of the large-$\Delta$/low density expansion. As
expected, the agreement is increasingly better for larger values of
the anisotropy.




\begin{thebibliography}{99}

\bibitem{Langen2015}
T. Langen, R. Geiger and J. Schmiedmayer, \emph{Ultracold Atoms Out of Equilibrium},
\href{https://doi.org/10.1146/annurev-conmatphys-031214-014548}{Ann. Rev. Cond. Matt. Phys. {\bf 6}, 201 (2015).}

\bibitem{Altman2015}
E. Altman, \emph{Nonequilibrium quantum dynamics in ultracold quantum
gases} in \emph{Strongly Interacting Quantum Systems out of Equilibrium:
Lecture Notes of the Les Houches Summer School}, eds T. Giamarchi,
A.J. Millis, O. Parcollet, H. Saleur and L.F. Cugliandolo,
\href{https://doi.org/10.1093/acprof:oso/9780198768166.003.0001}{\bf 99} (2012).

\bibitem{kww-06}
T. Kinoshita, T. Wenger,  D. S. Weiss, \emph{A quantum Newton's cradle},
\href{http://dx.doi.org/10.1038/nature04693}{Nature {\bf 440}, 900 (2006).}

\bibitem{GM:col_rev02} 
M. Greiner, O. Mandel, T.W. H\"ansch, and I. Bloch, \emph{Collapse and
revival of the matter wave field of a Bose-Einstein condensate},
\href{\doi10.1038/nature00968}{Nature {\bf 
419}, 51-54 (2002).}

\bibitem{hacker10}
L. Hackermuller, U. Schneider, M. Moreno-Cardoner, T. Kitagawa,
S. Will, T. Best, E. Demler, E. Altman, I. Bloch and B. Paredes,
\emph{Anomalous Expansion of Attractively Interacting Fermionic Atoms
  in an Optical Lattice},
\href{\doi10.1126/science.1184565}{Science {\bf 327}, 1621 (2010).}

\bibitem{tetal-11}
S. Trotzky Y.-A. Chen, A. Flesch, I. P. McCulloch, U. Schollw\"ock,
J. Eisert, and I. Bloch,  \emph{Probing the relaxation towards
equilibrium in an isolated strongly correlated 1D Bose
gas}, \href{\doi10.1038/nphys2232}{Nature Phys. {\bf 8}, 325 (2012).}

\bibitem{shr-12}
U. Schneider, L. Hackerm\"uller, J. P. Ronzheimer, S. Will, S. Braun,
T. Best, I. Bloch, E. Demler, S. Mandt, D. Rasch, and
A. Rosch, \emph{Fermionic transport and out-of-equilibrium dynamics in
a homogeneous Hubbard model with ultracold atoms},  
\href{\doi10.1038/nphys2205}{Nature Phys. {\bf 8}, 213 (2012).}

\bibitem{cetal-12}
M. Cheneau, P. Barmettler, D. Poletti, M. Endres, P. Schauss,
T. Fukuhara, C. Gross, I. Bloch, C. Kollath, and
S. Kuhr, \emph{Light-cone-like spreading of correlations in a quantum
many-body system},
\href{http://dx.doi.org/10.1038/nature10748}{Nature {\bf 481}, 484
(2012).}

\bibitem{langen13}
T. Langen, R. Geiger, M. Kuhnert, B. Rauer, and J. Schmiedmayer, \emph{Local emergence of thermal correlations in an isolated quantum many-body system}, 
\href{http://dx.doi.org/10.1038/nphys2739}{Nature Phys. {\bf 9}, 640 (2013).}

\bibitem{MM:Ising13} 
F. Meinert, M.J. Mark, E. Kirilov, K. Lauber, P. Weinmann, A.J. Daley,
and H.-C. N\"agerl, \emph{Quantum Quench in an Atomic One-Dimensional
Ising Chain},
\href{http://dx.doi.org/10.1103/PhysRevLett.111.053003}{Phys. Rev. Lett. 
{\bf 111}, 053003 (2013).}

\bibitem{ronzheimer13}
J.P. Ronzheimer, M. Schreiber, S. Braun, S.S. Hodgman, S. Langer,
I.P. McCulloch, F. Heidrich-Meisner, I. Bloch and U. Schneider,
\emph{Expansion dynamics of interacting bosons in homogeneous lattices
  in one and two dimensions},
\href{http://dx.doi.org/10.1103/PhysRevLett.110.205301}{Phys. Rev. Lett. {\bf
  110}, 205301 (2013).}

\bibitem{zoran1}
N. Navon, A.L. Gaunt, R.P. Smith and Z. Hadzibabic,
\emph{Critical Dynamics of Spontaneous Symmetry Breaking in a
  Homogeneous Bose gas},
\href{\doi10.1126/science.1258676}{Science {\bf 347}, 167 (2015).}

\bibitem{lev2017}
Y. Tang, W. Kao, K.-Y. Li, S. Seo, K. Mallayya, M. Rigol,
S. Gopalakrishnan and B.L. Lev, \emph{Thermalization near
integrability in dipolar quantum Newton's cradle}, 
\href{https://doi.org/10.1103/PhysRevX.8.021030}{Phys. Rev. X {\bf 8}, 021030 (2018).}

\bibitem{HLSI08} 
S. Hofferberth, I. Lesanovsky, T. Schumm, A. Imambekov, V. Gritsev, E. Demler, and J. Schmiedmayer, 
{\it Probing quantum and thermal noise in an interacting many-body system},
\href{http://dx.doi.org/10.1038/nphys941}{Nature Phys. {\bf 4}, 489 (2008)}.

\bibitem{KPIS10} 
T. Kitagawa, S. Pielawa, A. Imambekov, J. Schmiedmayer, V. Gritsev,
and E. Demler, {\it Ramsey Interference in One-Dimensional Systems: The
Full Distribution Function of Fringe Contrast as a Probe of Many-Body
Dynamics}, \href{http://dx.doi.org/10.1103/PhysRevLett.104.255302}{Phys. Rev. Lett. {\bf
    104}, 255302 (2010)}. 

\bibitem{KISD11} 
T. Kitagawa, A. Imambekov, J. Schmiedmayer, and E. Demler, 
{\it The dynamics and prethermalization of one-dimensional quantum
systems probed through the full distributions of quantum
noise,} \href{http://dx.doi.org/10.1088/1367-2630/13/7/073018}{New
J. Phys. {\bf 13}, 73018 (2011).} 

\bibitem{GKLK12} 
M. Gring, M. Kuhnert, T. Langen, T. Kitagawa, B. Rauer, M. Schreitl, I. Mazets, D. A. Smith, E. Demler, and J. Schmiedmayer, 
{\it Relaxation and Prethermalization in an Isolated Quantum System},
\href{http://dx.doi.org/10.1126/science.1224953}{Science {\bf 337}, 1318 (2012)}.

\bibitem{Greiner}
A. Mazurenko, C.S. Chiu, G. Ji, M.F. Parsons, M. Kanasz-Nagy,
R. Schmidt, F. Grusdt, E. Demler, D. Greif and M. Greiner, {\it
Experimental realization of a long-range antiferromagnet in the
Hubbard model with ultracold atoms}, \href{https://dx.doi.org/10.1038/nature22362}{Nature {\bf 545}, 462 (2017).}

\bibitem{Hill}
M. F\"orst et al, {\it Melting of Charge Stripes in Vibrationally
Driven $La_{1.875}Ba_{0.125}CuO_4$ 
: Assessing the Respective Roles of Electronic and Lattice Order in Frustrated Superconductors},
\href{https://doi.org/10.1103/PhysRevLett.112.157002}{Phys. Rev. Lett. {\bf 112}, 157002 (2014).}

\bibitem{Cavalieri}
R. Mankowsky, M. F\"orst and A. Cavalleri, {\it Non-equilibrium
control of complex solids by nonlinear phononics},
\href{https://dx.doi.org/10.1088/0034-4885/79/6/064503}{Rep. Progr. Phy. {\bf 79} 064503 (2016).}

\bibitem{Deutsch1991}
J. M. Deutsch, \emph{Quantum statistical mechanics in a closed
  system}, 
\href{http://dx.doi.org/10.1103/PhysRevA.43.2046}{Phys. Rev. A {\bf 43}, 2046
(1991).}

\bibitem{Srednicki1994}
M. Srednicki, \emph{Chaos and quantum thermalization},
\href{http://dx.doi.org/10.1103/PhysRevE.50.888}{Phys. Rev. E{\bf 50}, 888
(1994).}

\bibitem{Rigol2008}
M. Rigol, V. Dunjko, and 
M. Olshanii, \emph{Thermalization and its mechanism for generic
isolated quantum systems},
\href{\doi10.1038/nature06838}{Nature {\bf 452}, 854 (2008).}

\bibitem{Rigol2007}
M. Rigol, V. Dunjko, V. Yurovsky,  and M. Olshanii, \emph{Relaxation
in a Completely Integrable Many-Body Quantum System: An Ab Initio
Study of the Dynamics of the Highly Excited States of 1D Lattice
Hard-Core
Bosons}, \href{http://dx.doi.org/10.1103/PhysRevLett.98.050405}{Phys. Rev. Lett. {\bf
98}, 50405 (2007).}

\bibitem{Cassidy2011}
A.~C. Cassidy, C.~W. Clark, and M.~Rigol, \emph{Generalized
Thermalization in an Integrable Lattice
System}, \href{http://dx.doi.org/10.1103/PhysRevLett.106.140405}{Phys. Rev. Lett. {\bf
106}, 140405 (2011).}

\bibitem{Caux2013} 
J.-S.~Caux and F.H.L.~Essler, \emph{Time Evolution of Local
Observables After Quenching to an Integrable
Model}, \href{http://dx.doi.org/10.1103/PhysRevLett.110.257203}{Phys. Rev. Lett. {\bf
110}, 257203 (2013).}

\bibitem{Ilievski2015} E. Ilievski, J. De Nardis, B. Wouters,
J.-S. Caux, F.H.L. Essler, T. Prosen, \emph{Complete Generalized Gibbs
Ensemble in an interacting
Theory}, \href{http://dx.doi.org/10.1103/PhysRevLett.115.157201}{Phys. Rev. Lett. {\bf
115}, 157201 (2015).}


\bibitem{Pozsgay:13a}
B. Pozsgay, \emph{The generalized Gibbs ensemble for Heisenberg spin
chains}, \href{http://dx.doi.org/10.1088/1742-5468/2013/07/P07003}{J. Stat. Mech. P07003
(2013).}

\bibitem{FE_13b}
M.~Fagotti and F.H.L. Essler, \emph{Stationary behaviour of
observables after a quantum quench in the spin-1/2 Heisenberg XXZ
chain}, \href{http://dx.doi.org/10.1088/1742-5468/2013/07/P07012}{J. Stat. Mech. P07012
(2013).}

\bibitem{FCEC_14} M, Fagotti, M, Collura, F.H.L. Essler, and
P. Calabrese, \emph{Relaxation after quantum quenches in the spin-1/2
Heisenberg XXZ chain},
\href{http://dx.doi.org/10.1103/PhysRevB.89.125101}{Phys. Rev. B 
 {\bf 89}, 125101 (2014).}

\bibitem{neel_overlap}
B. Wouters, J. De Nardis, M. Brockmann, D. Fioretto, M. Rigol,
J.-S. Caux, {\it Quenching the Anisotropic Heisenberg Chain: Exact
Solution and Generalized Gibbs Ensemble Predictions},
\href{http://dx.doi.org/10.1103/PhysRevLett.113.117202}{Phys. Rev. Lett. {\bf
113}, 117202 (2014).}

\bibitem{QANeel} M. Brockmann, B. Wouters, D. Fioretto, J. De Nardis,
R. Vlijm and J.-S. Caux, \emph{Quench action approach for releasing
the N\'eel state into the spin-1/2 XXZ chain},
\href{http://dx.doi.org/10.1088/1742-5468/2014/12/P12009}{Stat. Mech.
P12009 (2014).}

\bibitem{XXZung} B. Pozsgay, M. Mesty\'an, M.A. Werner, M. Kormos,
  G. Zar\'and, and G. Tak\'acs, \emph{Correlations after Quantum
    Quenches in the XXZ Spin Chain: Failure of the Generalized Gibbs
    Ensemble}, \href{http://dx.doi.org/10.1103/PhysRevLett.113.117203}{Phys. Rev. Lett. {\bf
    113}, 117203 (2014).}

\bibitem{XXZunglong} M. Mesty\'an, B. Pozsgay, G. Tak\'acs, and
M.A. Werner, \emph{Quenching the XXZ spin chain: quench action
approach versus generalized Gibbs ensemble},
\href{http://dx.doi.org/10.1088/1742-5468/2015/04/P04001}{J. Stat. Mech. P04001
(2015). }

\bibitem{IQNB}
E. Ilievski, E. Quinn, J. De Nardis and M. Brockmann,
\emph{String-charge duality in integrable lattice models},
\href{http://dx.doi.org/10.1088/1742-5468/2016/06/063101}{J. Stat. Mech. 063101 (2016).}

\bibitem{iTEBD1}
G. Vidal, \emph{Classical Simulation of Infinite-Size Quantum Lattice
Systems in One Spatial Dimension},
\href{https://doi.org/10.1103/PhysRevLett.98.070201}{Phys. Rev. Lett. 
 {\bf 98}, 070201 (2007). }

\bibitem{iTEBD2}
R. Or\'us and G. Vidal, \emph{Infinite time-evolving block decimation
algorithm beyond unitary
evolution},\href{https://doi.org/10.1103/PhysRevB.78.155117}{ Phys. Rev. B {\bf 78},
155117 (2008). }

\bibitem{BPGDA09}
P. Barmettler, M. Punk, V. Gritsev, E. Demler, and E. Altman,
\emph{Relaxation of antiferromagnetic order in spin-1/2 chains following a quantum quench},
\href{https://doi.org/10.1103/PhysRevLett.102.130603}{Phys. Rev. Lett. 
 {\bf 102}, 130603 (2009)}.

\bibitem{BPGDA09b}
P. Barmettler, M. Punk, V. Gritsev, E. Demler, and E. Altman,
\emph{Quantum quenches in the anisotropic spin-1/2 Heisenberg chain:
different approaches to many-body dynamics far from equilibrium},
\href{https://doi.org/10.1088/1367-2630/12/5/055017}{New J. Phys.  {\bf 12}, 055017 (2010).}

\bibitem{SotiriadisCardy}
S. Sotiriadis and J. Cardy, {\it Quantum quench in interacting field
theory: A self-consistent
approximation}, \href{https://dx.doi.org/10.1103/PhysRevB.81.134305}{Phys. Rev. B
{\bf 81}, 134305 (2010).}

\bibitem{vNE2018}
Y.D. van Nieuwkerk and F.H.L. Essler, \emph{Self-consistent
time-dependent harmonic approximation for the Sine-Gordon model out of
equilibrium}, \href{http://arxiv.org/abs/1812.06690}{arXiv:1812.06690.}

\bibitem{Groha}
S. Groha, F.H.L. Essler and P. Calabrese, {\it Full Counting
Statistics in the Transverse Field Ising
Chain}, \href{https://doi.org//10.21468/SciPostPhys.4.6.043}{SciPost
Phys. 4, 043 (2018). }




\bibitem{EK:finiteT}
F.H.L. Essler and R.M. Konik, \emph{Finite Temperature Dynamical
  Correlations in Massive Integrable Quantum Field Theories},
\href{http://dx.doi.org/10.1088/1742-5468/2009/09/P09018}{J. Stat. Mech. P09018
  (2009).}

\bibitem{JEK}
A.J.A. James, F.H.L. Essler and R.M. Konik, \emph{Finite Temperature
Dynamical Structure Factor of Alternating Heisenberg Chains},
\href{https://dx.doi.org/10.1103/PhysRevB.78.094411}{Phys. Rev. B{\bf 78}, 094411 (2008).}

\bibitem{GKE}
W.D. Goetze, U. Karahasanovic and F.H.L. Essler, \emph{Low-Temperature
Dynamical Structure Factor of the Two-Leg Spin-1/2 Heisenberg Ladder},
\href{https://dx.doi.org/10.1103/PhysRevB.82.104417}{ Phys. Rev. B{\bf 82}, 104417 (2010).}

\bibitem{PT10}
B. Pozsgay and G. Takacs, \emph{Form factor expansion for thermal
correlators}, \href{https://dx.doi.org/10.1088/1742-5468/2010/11/P11012}{
J. Stat. Mech. P11012 (2010).}

\bibitem{ST12}
I.M. Szecsenyi and G. Takacs,
\emph{Spectral expansion for finite temperature two-point functions
and clustering}, \href{https://dx.doi.org/10.1088/1742-5468/2012/12/P12002}{J. Stat. Mech.P12002 (2012).}

\bibitem{CEF1} 
P. Calabrese, F.H.L. Essler, and M. Fagotti, \emph{Quantum Quench in
  the Transverse-Field Ising
  Chain}, \href{http://dx.doi.org/10.1103/PhysRevLett.106.227203}{Phys. Rev. Lett. {\bf
  106}, 227203 (2011).}

\bibitem{CEF2} 
P. Calabrese, F.H.L. Essler, and M. Fagotti, \emph{Quantum quench in
the transverse field Ising chain: I. Time evolution of order parameter
correlators}. \href{http://dx.doi.org/10.1088/1742-5468/2012/07/P07016}{J. Stat. Mech. P07016
(2012).}

\bibitem{SE:Ising} D. Schuricht and F.H.L. Essler, \emph{Dynamics in
the Ising field theory after a quantum quench}, 
\href{http://dx.doi.org/10.1088/1742-5468/2012/04/P04017}{J. Stat. Mech.,
P04017 (2012).}
\bibitem{sine-Gordon} 
B. Bertini, D. Schuricht, and F.H.L. Essler, \emph{Quantum quench in
the sine-Gordon
model}, \href{http://dx.doi.org/10.1088/1742-5468/2014/10/P10035}{J. Stat. Mech. P10035
(2014).}

\bibitem{SchriefferWolf}
A.~H.~MacDonald, S.~M.~Girvin, D.~Yoshioka, \href{https://doi.org/10.1103/PhysRevB.37.9753}{Phys. Rev. B {\bf 37}, 16 (1988).}

\bibitem{BEL}
L. Bonnes, F.H.L. Essler and A. L\"auchli, \emph{``Light-Cone''
Dynamics After Quantum Quenches in Spin Chains},
\href{http://dx.doi.org/10.1103/PhysRevLett.113.187203}{Phys. Rev. Lett. {\bf
113}, 187203 (2014). }

\bibitem{prethermal}
B. Bertini, F.H.L. Essler, S. Groha, N.J. Robinson, 
\emph{Thermalization and light cones in a model with weak integrability breaking},
\href{https://dx.doi.org/10.1103/PhysRevB.94.245117}{Phys. Rev. B {\bf 94}, 245117 (2016).}

\bibitem{Fagotti}
M. Fagotti and P. Calabrese, {\it Entanglement entropy of two disjoint
blocks in XY
chains}, \href{https://doi.org/10.1088/1742-5468/2010/04/P04016}{J. Stat. Mech. P04016
(2010)}.
 
\end{thebibliography}
\end{document}